\documentclass[a4paper,USenglish]{lipics-v2021}
\usepackage{amsmath,amssymb,mathtools,booktabs,array,tikz}
\usetikzlibrary{arrows.meta,calc,positioning,fit,backgrounds, shapes.geometric}
\hideLIPIcs
\author{Subrahmanyam Kalyanasundaram}
  {Department of CSE, IIT Hyderabad, India}
  {subruk@cse.iith.ac.in} %
  {} %
  {} %
\author{Subodh Kumar}
  {Department of CSE, IIT Hyderabad, India}
  {cs23resch11009@iith.ac.in} %
  {} %
  {} %
\authorrunning{S. Kalyanasundaram and S. Kumar}
\Copyright{Subrahmanyam Kalyanasundaram and Subodh Kumar}
\title{Conflict-Free Cuts: Exploring the Boundaries of
Tractability}
\titlerunning{Conflict-Free Cuts: Exploring the Boundaries of Tractability}
\AtBeginDocument{\hypersetup{pdfsubject={}}}
\newcommand{\CF}{\textnormal{\textsc{Conflict-Free Cut}}}
\newcommand{\EqSel}{\operatorname{DCS}}
\newcommand{\calC}{\mathcal C}
\newcommand{\bits}{\{0,1\}}
\newcommand{\YES}{\textnormal{\textsc{yes}}}
\newcommand{\NO}{\textnormal{\textsc{no}}}
\newcommand{\coc}{\operatorname{coc}}
\newcommand{\pw}{\operatorname{pw}}
\newcommand{\tw}{\operatorname{tw}}
\newcommand{\fvs}{\operatorname{fvs}}
\newcommand{\td}{\operatorname{td}}
\newcommand{\vi}{\operatorname{vi}}
\newcommand{\UG}{\operatorname{UG}}
\definecolor{ink}{HTML}{233444}
\definecolor{teal}{HTML}{147D92}
\definecolor{accent}{HTML}{AF473C}
\definecolor{gold}{HTML}{D7AA48}
\tikzset{vertex/.style={circle,draw=ink,fill=white,inner sep=1.5pt,minimum size=5mm},
  arr/.style={-{Stealth[length=2mm]},draw=ink},
  block/.style={draw=ink,rounded corners,align=center,inner sep=5pt},
  conflict/.style={draw=accent,dashed,thick}}

\definecolor{cfcHost}{HTML}{242834}
\definecolor{cfcConflict}{HTML}{B9362D}
\definecolor{cfcAdded}{HTML}{1769AA}
\definecolor{cfcFill}{HTML}{F1F3F5}
\definecolor{cfcBoundary}{HTML}{ABB3BC}
\tikzset{
 cfc edge/.style={draw=cfcHost,line width=.65pt},
 cfc conflict/.style={draw=cfcConflict,line width=.85pt,dash pattern=on 1.6pt off 1.1pt},
 cfc added/.style={draw=cfcAdded,line width=1.15pt},
 cfc port/.style={circle,fill=cfcHost,inner sep=1.1pt},
 cfc vertex/.style={circle,fill=cfcHost,inner sep=1.4pt},
 cfc panel/.style={anchor=west,font=\footnotesize\bfseries},
 cfc label/.style={font=\footnotesize}
}
\newcommand{\CFCompactBlock}[5]{%
 \begin{scope}[shift={(#2,#3)},xscale=#4]
  \coordinate (#1) at (0,0);
  \coordinate (#1-0) at (-.73,0);
  \coordinate (#1-1) at (-.365,.632);
  \coordinate (#1-2) at (.365,.632);
  \coordinate (#1-3) at (.73,0);
  \coordinate (#1-4) at (.365,-.632);
  \coordinate (#1-5) at (-.365,-.632);
  \filldraw[fill=cfcFill,draw=cfcBoundary,line width=.45pt]
   (#1-0)--(#1-1)--(#1-2)--(#1-3)--(#1-4)--(#1-5)--cycle;
  \foreach \j in {0,...,5}{
   \node[cfc port] at (#1-\j) {};
   \node[font=\scriptsize,inner sep=0pt] at ($(#1)!.72!(#1-\j)$) {$\j$};
  }
  \node[font=\small,inner sep=0pt] at (0,-.02) {#5};
 \end{scope}
}
\newcommand{\CFCompactEquality}[4]{%
 \coordinate (#1) at (#2);
 \draw[cfc edge] (#3-2)--(#1)--(#3-3);
 \draw[cfc edge] (#4-0)--(#1)--(#4-1);
 \draw[cfc conflict] ($(#1)!.34!(#3-2)$) to[bend right=25] ($(#1)!.34!(#3-3)$);
 \draw[cfc conflict] ($(#1)!.34!(#4-0)$) to[bend left=25] ($(#1)!.34!(#4-1)$);
 \node[cfc vertex] at (#1) {};
}

\begin{document}
\nolinenumbers
\maketitle
\begin{abstract}
A conflict-free cut of a host graph $G$ is an edge cut containing no conflicting pair of edges. The conflicts are specified by a conflict graph $\widehat G$ whose vertex set is $E(G)$. The  conflict-free cut problem asks if the given instance $(G, \widehat G)$ has a conflict-free cut. When restricted to planar $G$ and 1-regular $\widehat G$, the problem is polynomial time solvable when $G$ has maximum degree 4, but NP-complete already at maximum degree 5. We study the complexity of conflict-free cut problem under restrictions on the degrees of $G$, the components of $\widehat G$, and the locality of the conflicts. A conflict is local if its two edges share an endpoint in $G$. 

We prove the following results.
\begin{itemize}
\item If $\Delta(\widehat G)\le1$, the problem is polynomial-time solvable when $|E(G)|\le2|V(G)|$, and hence on all graphs of maximum degree four. 
The algorithm uses a directed-cycle formulation that
also underlies our surplus algorithms. %
This extends the known algorithm for planar $4$-regular graphs $G$ with 1-regular $\widehat G$. In contrast, the problem is NP-complete on $5$-regular graphs with exactly $1$-regular conflicts.
\item For $\Delta(\widehat G)\le1$, we give a constructive fixed-parameter algorithm in the degree-four surplus $s_4(G)=\sum_v\max\{0,d_G(v)-4\}$. For planar graphs, its running time has a single exponential dependence on the degree-four surplus parameter.
For arbitrary conflicts, we give an XP algorithm in the conflict surplus $\eta=\max\{0,|E(\widehat G)|-|V(G)|\}$.
\item The problem is NP-complete on $4$-regular hosts
with local exactly $2$-regular conflicts, even when
the conflict components are triangles and four-cycles.
For local conflicts on hosts of maximum degree four,
conflict graphs consisting only of pairs or only of
triangles each admit polynomial-time algorithms,
but allowing both component types gives NP-completeness.
The triangle algorithm combines forced monochromatic
vertex sets with majority propagation. More generally, for hosts of maximum degree four with local conflicts, we give a
complete classification when the conflict graph
is a disjoint union of cliques.

\item An $a$-component-order deletion set, or $a$-COC,
is a vertex set whose deletion leaves components
of order at most $a$.
For local conflicts, the problem is FPT parameterized
by the minimum size of a $2$-COC, but is NP-complete
even with a supplied $3$-COC of size six.
Without locality, NP-completeness already holds
with a supplied $2$-COC of size six.
Both hardness results hold even for exactly
$1$-regular conflicts.

\end{itemize}
\end{abstract}
\clearpage
\label{page:mainstart}
\section{Introduction}\label{sec:introduction}
We study the conflict-free cut problem under restrictions on the degrees of the input graph and on the structure of the conflicts. The problem is defined as follows.

\begin{definition}[Conflict-Free Cut]\label{def:cfcut}
Given a simple connected graph $G=(V,E)$ and a conflict graph $\widehat G=(E,\calC)$, the \emph{Conflict-Free Cut} (\emph{CF-cut}) problem asks whether there is a set $F\subseteq E$ such that $G-F$ is disconnected and $F$ is independent in $\widehat G$.
\end{definition}

We call $G$ the \emph{host graph}, or simply the \emph{host}, and $\widehat G$ the \emph{conflict graph}. The vertices of $\widehat G$ are the edges of $G$. 
Two host edges conflict when they are adjacent
in $\widehat G$.
Throughout, a cut means an edge cut.

Conflict constraints have been studied for several optimization problems, including spanning trees~\cite{Barros2022,Darmann2011}, knapsack~\cite{Pferschy2009,Bettinelli2017}, bin packing~\cite{Capua2018}, matchings and paths~\cite{agrawal2020parameterized}, and feedback vertex set~\cite{conflict-FVS}. In each of these problems, certain pairs of objects are forbidden from appearing together in a solution. Rauch, Rautenbach, and Souza~\cite{RRS} introduced this restriction for cuts. They showed that the CF-cut problem is NP-complete even when $G$ has maximum degree five and $\widehat G$ is $1$-regular. In this case, every edge of $G$ conflicts with exactly one other edge. They also gave a fixed-parameter algorithm parameterized by the vertex-cover number of $G$, and studied restrictions on both $G$ and $\widehat G$.

The \textsc{Matching Cut} problem is a special case of CF-cut. A matching cut is a cut whose edges form a matching in $G$. Setting $\widehat G=L(G)$, the line graph of $G$, makes every two incident edges conflict and gives precisely this problem. \textsc{Matching Cut} is NP-complete in general~\cite{Chvatal1984}, including on planar graphs~\cite{bonsma2009complexity}, but several restrictions make it tractable. 
For example, it is polynomial-time solvable on graphs
of bounded treewidth~\cite{GomesSau2019}, as well as on
claw-free and $P_6$-free graphs; more general results
are known for $H$-free graph classes~\cite{LPR2022,LPR2023}.

One feature of \textsc{Matching Cut} is that its conflicts are local: conflicting edges always share an endpoint in $G$. This motivates the following terminology. We say that a conflict graph is \emph{local} if every conflict is between two edges of $G$ that share an endpoint. We say that the conflicts are \emph{matching} if $\Delta(\widehat G)\le1$; this permits an edge of $G$ to have no conflicting partner. When we require every edge of $G$ to participate in exactly $r$ conflicts, we say that $\widehat G$ is \emph{exactly $r$-regular}. The two restrictions, locality and matching conflicts, are independent. %

\subsection{Our contributions and discussion}

Kalyanasundaram and Kumar~\cite{KK} showed that planar
$4$-regular hosts with exactly $1$-regular conflicts
always admit a CF-cut unless the host is the octahedron
graph. Lemma~\ref{lem:U} gives an instance on the
octahedron with no CF-cut. They asked whether
this tractability extends to all $4$-regular graphs.
We answer this question affirmatively through a directed-cycle
formulation, and use this formulation to investigate how far
tractability persists beyond degree four. We then study the
effect of allowing more conflicts per edge, and the role of
conflict locality under structural restrictions on the host.
Throughout this discussion, $n=|V(G)|$, $m=|E(G)|$, and
$q=|E(\widehat G)|$.

We first prove that CF-cut is polynomial-time solvable for
matching conflicts whenever $m\le2n$
(Theorem~\ref{thm:density}). This includes all hosts of maximum
degree at most four. When $m<2n$, the counting argument of
Rauch et al.~\cite{RRS} already guarantees a conflict-free
cut separating a single vertex. In the remaining case, if
no such cut works, the conflict pairs partition the edges
and determine an orientation with indegree two at every
vertex. We show that a CF-cut exists precisely when this
digraph contains two vertex-disjoint directed cycles.
McCuaig's algorithm~\cite{McCuaig1993} then decides the
instance in polynomial time. 
Unlike the planar case, infinitely many uncuttable
$4$-regular instances exist~\cite{RRS}.

At degree five, we strengthen the known
hardness~\cite{RRS,KK} in two directions:
NP-completeness holds on exactly $5$-regular hosts
(Theorem~\ref{thm:degree-five}(ii)) and on hosts of
maximum degree five with local conflicts
(Theorem~\ref{thm:degree-five}(i)).
Both results require exactly $1$-regular conflicts.
These results motivate measuring the extent to which the
host exceeds degree four. Define the \emph{degree-four
surplus} by
\(
 s_4(G)=\sum_{v\in V(G)}\max\{0,d_G(v)-4\}.
\)
For matching conflicts, we give a constructive FPT
algorithm in $s_4(G)$ (Theorem~\ref{thm:s4}), with
running time $2^{O(s_4(G))}n^{O(1)}$ for planar hosts
(Theorem~\ref{thm:planar-s4}).
Neither algorithm requires locality.
When $\Delta(G)\le5$, $s_4(G)$ counts the degree-five
vertices, giving tractability when their number is
bounded despite hardness for the case  when $\Delta(G) \leq 5$.

We identify a second measure of
departure from the tractable case. Once all single-vertex
cuts have been ruled out, choose one conflict centered at
each vertex. The \emph{conflict surplus}
$\eta=\max\{0,q-n\}$ counts the unchosen conflicts.
We obtain an XP algorithm parameterized by $\eta$, allowing
arbitrary conflicts (Theorem~\ref{thm:eta}).
Both surplus algorithms extend the degree-four
cycle formulation. The XP algorithm uses equality
choices and bounded-size directed patterns.
For FPT, matching conflicts let $s_4(G)$ bound
both the number of additional conflicts and the
auxiliary digraph's positive degree imbalance.
A large cycle packing yields a cut; otherwise,
a bounded directed feedback vertex set and this
imbalance bound give bounded pathwidth for
dynamic programming.
After the single-vertex test, matching instances
satisfy $4\eta\le s_4(G)$, but bounded $\eta$ does
not bound $s_4(G)$. FPT in $\eta$ remains open,
even for exactly $1$-regular conflicts.

Returning to exactly $4$-regular hosts, allowing
exactly two conflicts per edge makes the problem
NP-complete even under locality, with $\widehat G$
consisting only of triangles and four-cycles
(Theorem~\ref{thm:two-four}).
Together with the matching-conflict algorithm,
this gives a sharp conflict-degree boundary.

A second tractable structure arises for local
conflicts whose components are triangles or isolated
vertices, giving a polynomial-time algorithm at
maximum degree four (Theorem~\ref{thm:triangles}).
In a binary coloring representing a cut, each conflict
triangle forces either a monochromatic host triangle
or a majority equation at a star center.
Merge intersecting triangles of the first type into
classes. If no single-vertex cut works, a CF-cut
exists exactly when there are at least two classes;
majority propagation colors the remaining vertices.
Consequently, conflict graphs consisting of
disjoint edges and those consisting of disjoint triangles
each yield tractable cases, but allowing both types
together gives NP-completeness.

We extend this observation to a complete classification
for local cluster conflict graphs
(Theorem~\ref{thm:component-dichotomy}).
The algorithms also permit
controlled departures from the tractable component
families. In particular, for local conflict graphs with
components in $\{K_1,K_2,K_3\}$, we obtain FPT in
$s_4(G)+\min\{t_2,t_3\}$, where $t_2$ and $t_3$ count
edge and triangle components, respectively
(Corollary~\ref{cor:component-fpt} (ii)).
The majority algorithm also admits a single-exponential
extension parameterized by degree surplus and the number
of host vertices involved in exceptional conflict
components (Theorem~\ref{thm:exceptions}).

Finally, motivated by the structural tractability of
\textsc{Matching Cut}, we ask when locality makes CF-cut
tractable without restricting host degrees or conflict
components. Even simple host classes remain difficult:
CF-cut is NP-complete on cliques with local matching
conflicts, allowing unpaired edges
(Proposition~\ref{prop:clique}), and with a universal
vertex even under local exactly $1$-regular conflicts
(Theorem~\ref{thm:domination}).
In particular, locality does not yield tractability
parameterized by distance to cluster graphs alone.

We obtain a  positive result when the remaining components have order at most two. 
An \emph{$a$-component-order deletion
set}, or \emph{$a$-COC}, is a vertex set whose deletion
leaves components of order at most $a$; its minimum size
is denoted by $\coc_a(G)$. The case $a=1$ is vertex cover.
For local conflicts, we obtain an $O^*(2^k)$ algorithm
given a $2$-COC of size $k$, and an $O^*(3^k)$ algorithm
including the search for such a set
(Theorem~\ref{thm:scope}).
Every local conflict involves three connected host
vertices, so fixing the colors of a $2$-COC leaves
at most two unknown colors per constraint,
reducing the remaining problem to 2-SAT.
The algorithm allows arbitrarily many conflicts
per edge.

Both locality and the residual component size are
essential. With local exactly $1$-regular conflicts,
CF-cut is NP-complete even with a supplied $3$-COC of
size six (Theorem~\ref{thm:coc-hardness}(i)); the same construction
bounds treewidth and feedback-vertex number by constants.
Without locality, hardness already holds with a supplied
$2$-COC of size six and exactly $1$-regular conflicts
(Theorem~\ref{thm:coc-hardness}(ii)).
Allowing multiple conflicts per edge reduces the
local $3$-COC hardness bound to two, whereas a
supplied singleton $a$-COC permits polynomial time
for every fixed $a$, even without locality
(Appendix~\ref{app:twohubs}).

\begin{table}[tb]
\caption{A summary of the main results. The conflicts
are specified by $\widehat G$. ``Matching'' permits
unpaired edges, whereas ``exactly $r$-regular'' does
not. The component-count algorithms are summarized
separately in Table~\ref{tab:component-parameters}.}
\label{tab:main}
\centering
\begin{tabular}{
>{\raggedright\arraybackslash}p{.27\linewidth}
>{\raggedright\arraybackslash}p{.30\linewidth}
>{\raggedright\arraybackslash}p{.32\linewidth}}
\toprule
Restriction on $G$ & Restriction on conflicts & Complexity\\
\midrule

$m\le2n$
& Matching
& Polynomial (Theorem~\ref{thm:density})\\

$5$-regular
& Exactly $1$-regular
& NP-complete (Theorem~\ref{thm:degree-five}(ii))\\

Maximum degree five
& Local, exactly $1$-regular
& NP-complete (Theorem~\ref{thm:degree-five}(i))\\

\addlinespace

Param. degree-four surplus $s_4(G)$
& Matching
& FPT; planar $2^{O(s_4)}n^{O(1)}$
(Theorems~\ref{thm:s4},~\ref{thm:planar-s4})\\

Param. conflict surplus $\eta$
& Arbitrary
& XP (Theorem~\ref{thm:eta})\\

\addlinespace

$4$-regular
& Local, exactly $2$-regular; components $C_3,C_4$
& NP-complete (Theorem~\ref{thm:two-four})\\

Maximum degree four
& Local; conflict graph is a cluster graph
& Classification by allowed components
(Theorem~\ref{thm:component-dichotomy})\\

\addlinespace

Parameter $\coc_2(G)$
& Local
& $O^*(3^{\coc_2(G)})$ (Theorem~\ref{thm:scope})\\

$\coc_3(G)\le6$
& Local, exactly $1$-regular
& NP-complete (Theorem~\ref{thm:coc-hardness}(i))\\

$\coc_2(G)\le6$
& Exactly $1$-regular
& NP-complete (Theorem~\ref{thm:coc-hardness}(ii))\\

\bottomrule
\end{tabular}
\end{table}

\subsection{Related results}
Rauch et al.~\cite{RRS} showed that CF-cut is FPT in
the treewidth of a graph on $V(G)\cup E(G)$ encoding
incidence and conflicts; this parameter differs
from the treewidth of $G$ alone. They also proved
parameterized hardness for the clique-cover number
of $\widehat G$, even on hosts with feedback-vertex
number one.

When $\widehat G$ is a cluster graph (a disjoint union
of cliques), coloring host edges by their conflict
components identifies CF-cuts with rainbow cuts.
Huang and Li~\cite{HuangLi} study rainbow disconnection,
requiring a rainbow cut for every vertex pair;
we seek one nontrivial cut for the prescribed coloring.
Cluster conflicts also define a partition matroid,
connecting this case to the separation algorithms
of Banik et al.~\cite{BanikEtAl}, parameterized by
solution size.

Full proofs of statements marked with $\star$,
together with supplementary results, appear
in the appendices.

\section{Preliminaries}\label{sec:preliminaries}

Graphs are finite, simple, and undirected unless
stated otherwise; hosts are connected and have
at least two vertices.
We use standard terminology~\cite{diestel2005graph}.
For $G=(V,E)$, write $n=|V|$, $m=|E|$,
$uv=\{u,v\}$, and $d_G(v)$ for vertex degree.
Write $\Delta(G)$ for maximum degree and $G-X$
for vertex deletion.
The graphs $K_a,P_a,C_a$ are the complete graph,
path, and cycle on $a$ vertices, respectively.

\subparagraph*{Conflicts and colorings.}
The conflict graph is  denoted by $\widehat G=(E,\calC)$, and $q=|\calC|$ is the number of conflict pairs. 
An isolated vertex of $\widehat G$ corresponds to an \emph{unpaired edge} of $G$. 
An instance $(G,\widehat G)$ with no CF-cut is called \emph{uncuttable}. 
For $\emptyset \ne S\subsetneq V$, let
\(
 \delta_G(S)=\{uv\in E:|\{u,v\}\cap S|=1\}.
\)
It suffices to consider cuts of this form: if $F$ is
a CF-cut, then any component $S$ of $G-F$ satisfies
$\delta_G(S)\subseteq F$, so $\delta_G(S)$ is also
conflict-free.

We may represent a cut by a coloring $f:V\to\bits$, assigning one color to each side. An edge $uv$ \emph{crosses} the cut when $f(u)\ne f(v)$, and is \emph{monochromatic} otherwise. A coloring is \emph{feasible} if it satisfies every conflict; it need not use both colors. Thus a solution to CF-cut is a nonconstant feasible coloring. For a conflict $\{ab,cd\}$, feasibility means
 $(f(a)=f(b))\lor(f(c)=f(d))$.
Both constant colorings are feasible. Complementing all colors preserves feasibility.

For a local conflict $\{xy,yz\}$, call $y$ its \emph{center} and $\{x,y,z\}$ its \emph{scope}. Define the Boolean relation $R$ as follows: 
\begin{equation}\label{eq:R}
 R=\{(a,b,c)\in\bits^3:a=b\text{ or }b=c\}=\bits^3\setminus\{010,101\}.
\end{equation}
We write $R(a,b,c)$ to mean that $(a,b,c)\in R$. The local conflict is therefore the condition $R(f(x),f(y),f(z))$. %

\subparagraph*{Digraphs and running times.}
For a digraph $D$, write $A(D)$ for its arc set
and $d_D^-(v),d_D^+(v)$ for indegree and outdegree.
The underlying graph $\UG(D)$ forgets arc
directions and merges opposite arcs; its components
are the weak components of $D$.
Auxiliary digraphs may have opposite arcs.
Disjoint directed cycles are vertex-disjoint
unless stated otherwise.

Since $m=O(n^2)$ and $q=O(m^2)$, the input size is
polynomial in $n$; we therefore write $n^{O(1)}$
for polynomial factors. An algorithm is fixed-parameter tractable (FPT) in $k$ if it runs in $F(k)n^{O(1)}$ time for a computable function $F$, with exponent independent of $k$. An XP algorithm runs in $F(k)n^{g(k)}$ time for computable functions $F,g$. The notation $O^*(\cdot)$ suppresses polynomial factors in the input size. We refer to Cygan et al.~\cite{Cygan2015} for parameterized-complexity terminology. Our constructive algorithms return a cut whenever the answer is yes.

\section{The boundary degrees four and five}\label{sec:cycles}
For matching conflicts, testing single-vertex cuts
reveals a directed-cycle formulation when $m\le2n$.
This formulation also underlies the surplus algorithms.

\begin{theorem}\label{thm:density}
CF-cut is polynomial-time solvable for matching
conflicts when $m\le2n$, and hence on all hosts
of maximum degree four.
\end{theorem}

We need the following observation, which  also applies when conflicts are not matching. It is the starting point for the surplus algorithms in the next section. %

\begin{observation}[Single-vertex Cuts] \label{lem:singleton}
If no cut $\delta_G(\{v\})$ is conflict-free, then every vertex is the center of a conflict and $q\ge n$. If the conflicts form a matching, then $m\ge2n$.
\end{observation}
\begin{proof}
For each vertex $v$, choose a conflicting pair contained in $\delta_G(\{v\})$. Since $G$ is simple, two distinct edges share at most one endpoint. Hence the pairs chosen at different vertices are distinct, giving $q\ge n$. If the conflicts form a matching, no edge appears in two of the chosen pairs. Their $2n$ edges are therefore distinct.
\end{proof}

We use the following algorithmic consequence of
McCuaig's characterization~\cite[Section~9]{McCuaig1993}. %

\begin{theorem}[{McCuaig~\cite[Section~9]{McCuaig1993}}]\label{ext:mccuaig}
Given a digraph $D$, one can find two vertex-disjoint directed cycles, or determine that none exist, in time polynomial in $|V(D)|+|A(D)|$.
\end{theorem}

\begin{proof}[Proof of Theorem~\ref{thm:density}]
Test the single-vertex cuts $\delta_G(\{v\})$ for all vertices $v$. If none of these are CF-cuts, Observation~\ref{lem:singleton} implies $m \geq 2n$. Together with the condition $m\leq 2n$ in the statement, we have $m=2n$ and $q=n$. The $n$ conflicting pairs use $2n$ distinct edges, so they use every edge of $G$. Since the conflicts form a matching, there are no other conflicts and no unpaired edges. Each chosen pair is local, and precisely one pair is centered at each vertex.

Orient each edge towards the center of its pair to obtain the directed graph $D$, which has indegree two at every
vertex. 
We show that $G$ has a CF-cut if and only if $D$ has two vertex-disjoint directed cycles. Suppose \(D\) has two vertex-disjoint directed cycles. Retain the arcs of these cycles and choose one arbitrary incoming arc at every vertex outside them. Let \(F\) be the resulting spanning subdigraph. Every vertex of \(F\) has indegree one, so each weak component consists of exactly one directed cycle with trees directed away from it. The two retained cycles therefore belong to different components, and \(\UG(F)\) is disconnected.

Let \(S\) be the vertex set of one component of \(\UG(F)\), and consider the cut \(\delta_G(S)\). No edge represented by an arc of \(F\) crosses this cut. At each vertex \(v\), the selected incoming arc represents one of the two edges of the conflict centered at \(v\). Thus every conflict has a noncrossing edge, so \(\delta_G(S)\) is conflict-free.

Conversely, a CF-cut leaves every vertex with an
in-neighbor on the same side. Since the host graph is finite, repeatedly following
such predecessors must revisit a vertex, producing
a directed cycle within that side. Applying this
argument on both sides gives two vertex-disjoint
directed cycles.

We can use Theorem~\ref{ext:mccuaig} to decide whether the disjoint directed cycles exist in $D$.
\end{proof}

\begin{remark}\label{rem:planar-obstruction}
Kalyanasundaram and Kumar~\cite{KK} proved that
planar $4$-regular hosts with exactly $1$-regular
conflicts admit a CF-cut unless the host is the
octahedron.
The same holds for matching conflicts by
singleton preprocessing.
General $4$-regular hosts have infinitely many
uncuttable instances~\cite{RRS};
Theorem~\ref{thm:density} decides them without
classifying their orientations.

McCuaig's characterization \cite{McCuaig1993} also lets one derive an alternate proof for the fact that the octahedron is the lone uncuttable planar 4-regular graph with matching conflicts.
\end{remark}

The degree-four bound is tight. The following theorem
establishes hardness at degree five, both under locality
and under exact regularity of the host.

\begin{theorem}[$\star$]\label{thm:degree-five}
\CF{} is NP-complete on simple connected hosts with
exactly $1$-regular conflicts in each of the following cases:
\begin{enumerate}[(i)]
\item $\Delta(G)= 5$ and every conflict is local;
\item $G$ is $5$-regular, without a locality requirement.
\end{enumerate}
\end{theorem}

\subparagraph*{Proof overview.}
Both reductions use octahedra forced monochromatic
to represent variable occurrences.
Connectors enforce agreement and the source
constraints while keeping conflicts local and
maximum degree five.
For part~(ii), a modified construction completes
the host to degree five using edges forced
monochromatic, which can be paired without
changing feasible colorings.
The full reductions appear in
Appendices~\ref{app:localfiveproof}
and~\ref{app:exactfive}.

The boundary between degrees four and five motivates
the surplus algorithms in the next section: we measure
how far the host exceeds degree four and ask whether
this departure can be confined to a parameter.

\section{Surplus algorithms}\label{sec:surplus}

We extend the directed-cycle formulation by retaining
one conflict centered at each vertex and replacing
the remaining conflicts by equality choices.
The resulting selector problem gives XP in $\eta$
through bounded-size directed patterns.
For matching conflicts, we obtain FPT in $s_4(G)$
by combining directed-cycle packing, degree imbalance,
and dynamic programming on a path decomposition.

\subsection{Selectors and the Conflict-surplus algorithm}\label{sec:conflictsurplus}

\begin{definition}[Disjoint cycles and selectors]
\label{def:eqsel}
Let $D$ be a loopless digraph in which each vertex $v$
has exactly two distinct in-neighbors $p_v,q_v$,
and let $Q\subseteq\binom{V(D)}2$ be a set of
\emph{equality pairs}.
A \emph{selector} is a spanning subdigraph of $D$
with exactly one incoming arc at each vertex.
The problem \emph{Disjoint Cycles and Selectors},
denoted $\EqSel(D,Q)$, asks whether a selector $F$
exists such that $\UG(F)+Q$ is disconnected,
where the pairs in $Q$ are added as undirected edges.
\end{definition}

The selector formulation has the following
coloring interpretation.

\begin{lemma}\label{lem:selector}
Every weak component of a selector consists of one directed cycle and (possibly) trees directed away from that cycle. Further, $\EqSel(D,Q)$ has a solution if and only if there is a
nonconstant coloring $f:V(D)\to\bits$ such that
\[
 R(f(p_v),f(v),f(q_v))\quad\text{for every }v\in V(D),
 \qquad f(a)=f(b)\quad\text{for every }ab\in Q.
\]
\end{lemma}

\begin{proof}
Following chosen predecessors from any vertex of a
selector $F$ reaches a directed cycle. The endpoints
of each selected arc reach the same cycle, so each
weak component consists of one directed cycle and
trees directed away from it. This includes cycles
formed by two opposite arcs.

If $\UG(F)+Q$ is disconnected, color one component
one and all others zero. Every vertex agrees with
its chosen predecessor, and every equality pair
has equally colored endpoints. This gives the
required nonconstant coloring.

Conversely, given such a coloring, choose at each
vertex an incoming arc from a vertex of the same
color; its $R$-constraint (defined in Equation (\ref{eq:R}))  guarantees this choice.
The resulting selector and all equality edges are
monochromatic. Since the coloring is nonconstant,
$\UG(F)+Q$ is disconnected.
\end{proof}

\begin{lemma}\label{lem:many}
Suppose $|Q| = h$. If $D$ has $h+2$ vertex-disjoint directed cycles, then $\EqSel(D,Q)$ has a solution. 
\end{lemma}
\begin{proof}
Extend the given cycles to a selector $F$ by choosing
an arbitrary incoming arc at every other vertex.
By Lemma~\ref{lem:selector}, the given cycles lie in
distinct weak components, so $F$ has at least $h+2$
components. Adding $h$ equality edges reduces this
number by at most $h$, leaving $\UG(F)+Q$ disconnected.
\end{proof}

Figure~\ref{fig:selectors} in Appendix~\ref{app:xp}
illustrates the selector structure and the
addition of equality edges.

Assume all single-vertex cuts fail. Choose one
centered conflict $P_v=\{vp_v,vq_v\}$ at each vertex
$v$ and insert the arcs $p_v\to v,q_v\to v$ into $D$.
These are $n$ distinct conflicts, giving a digraph
with $n$ vertices and $2n$ arcs.
A host edge chosen at both endpoints contributes
both opposite arcs, which must be retained to
represent the two chosen conflicts.

For each of the remaining $\eta=q-n$ conflicts
$\{ab,cd\}$, choose either $f(a)=f(b)$ or $f(c)=f(d)$.
For each set of choices, let $Q$ contain the
corresponding equality pairs, removing duplicates.
This gives at most $2^\eta$ instances of
$\EqSel(D,Q)$, each with $h=|Q|\le\eta$.

\begin{lemma}\label{lem:branch}
The original instance has a CF-cut if and only if
one of the constructed instances $\EqSel(D,Q)$
has a solution.
\end{lemma}
\begin{proof}
Let $f$ be a nonconstant feasible coloring of the
original instance. For each unchosen conflict,
select an equality satisfied by $f$.
Then $f$ satisfies all predecessor constraints of
$D$ and all equalities in $Q$, so
Lemma~\ref{lem:selector} gives a solution to
$\EqSel(D,Q)$.

Conversely, a solution to a constructed $\EqSel(D,Q)$ instance
gives, by Lemma~\ref{lem:selector}, a nonconstant
coloring satisfying all predecessor constraints
and equalities in $Q$. These satisfy every chosen
conflict through its predecessor constraint and
every unchosen conflict through its selected
equality. Hence the coloring defines a CF-cut.
\end{proof}

The equality-branch reduction yields the following
algorithm for arbitrary conflicts.

\begin{theorem}[$\star$]\label{thm:eta}
For arbitrary conflicts, \CF{} is in XP parameterized by $\eta=\max\{0,q-n\}$. More precisely, it is decidable in $a(\eta)n^{c(\eta)}$ time for computable functions $a,c$. In particular, $q\le n$ is polynomial-time solvable, and $q<n$ admits a single-vertex cut.
\end{theorem}

\subparagraph*{Proof overview.}
For a branch with $h=|Q|$, $h+2$ disjoint
directed cycles give a solution by
Lemma~\ref{lem:many}.
Otherwise, a witnessing selector has at most
$h+1$ components.
Retain its cycles and paths to equality endpoints,
then prune other trees and suppress unnecessary
nonterminals. The resulting pattern has at most
$7h+1$ vertices and remains disconnected after
adding the equalities.

Algorithms of Reed et al.~\cite{RRST} test
the cycle-packing condition and whether each
such pattern occurs with equality endpoints
fixed. The number of patterns and polynomial
exponents depend only on $h$.
Applying these tests to at most $2^\eta$
branches with $h\le\eta$ gives XP.
Appendix~\ref{app:xp} supplies the formal
definitions, external results, and complete proof.

\subsection{An FPT algorithm for degree-four surplus}
We now use degree imbalance to strengthen the
selector approach to an FPT algorithm for
matching conflicts.

A \emph{path decomposition} of a graph $G$ is a
sequence of vertex bags such that every edge has
both endpoints in some bag and each vertex occurs
in a nonempty interval of bags.
Its \emph{width} is the largest bag size minus one.
The \emph{pathwidth} $\pw(G)$ is the minimum width
of a path decomposition of $G$.

\begin{theorem}\label{thm:s4}
For matching conflicts, \CF{} is FPT parameterized by $s_4(G)=\sum_v\max\{0,d_G(v)-4\}$. A cut can be produced in the same parameterized time. Consequently, on graphs of maximum degree at most five, finding a cut is FPT in the number of degree-five vertices.
\end{theorem}

A \emph{directed feedback vertex set} of a digraph
$D$ is a set $Z\subseteq V(D)$ such that $D-Z$
contains no directed cycle.
We use the following constructive alternative.

\begin{lemma}[$\star$]\label{lem:constructive-packing}
There is a computable function $R$ and an algorithm
that, given a digraph $D$ and an integer $t\ge1$,
returns either $t$ vertex-disjoint directed cycles
or a directed feedback vertex set of size at most
$R(t)$, in FPT time parameterized by $t$.
\end{lemma}

The proof combines directed-cycle packing,
the directed Erd\H{o}s--P\'osa theorem, and an
FPT algorithm for directed feedback vertex set;
the external statements and derivation appear
in Appendix~\ref{app:fpt-packing}.

A bounded feedback vertex set alone does not
bound the pathwidth needed by our algorithm.
For any digraph $H$, define its positive degree
imbalance by
$\iota(H)=\sum_{v\in V(H)}
\max\{0,d_H^+(v)-d_H^-(v)\}$.
Since $D$ has indegree two,
$\iota(D)=\sum_{v\in V(D)}\max\{0,d_D^+(v)-2\}$.

We adapt the degree-balance argument of
Gutin et al.~\cite[Lemmas~2--3]{GJSW}, which uses
Eulerian balance and feedback arc deletion to
bound cutwidth. Here positive imbalance and
feedback vertex deletion give a pathwidth bound.

\begin{lemma}\label{lem:imbalance}
If $D$ has indegree two and $Z$ is a directed feedback vertex set of size $r$, then a path decomposition of $\UG(D)$ of width at most $\iota(D)+3r$ can be constructed in polynomial time from $Z$.
\end{lemma}
\begin{proof}
Let $D_0=D-Z$ and write
$b_H(v)=d_H^+(v)-d_H^-(v)$.
For vertex sets $X,Y\subseteq V(H)$, let
$A_H(X,Y)$ denote the arcs directed from $X$
to $Y$.

For $v\notin Z$, deleting outgoing arcs to $Z$
cannot increase its positive imbalance, while
deleting incoming arcs from $Z$ increases it
by at most $|A_D(Z,\{v\})|$.
Since
$d_D^+(z)=2+b_D(z)\le2+\max\{0,b_D(z)\}$,
we obtain
\(
\iota(D_0)
\le \sum_{v\notin Z}\max\{0,b_D(v)\}
   +\sum_{z\in Z}d_D^+(z)
\le \iota(D)+2r.
\)

Take a topological ordering $v_1,\ldots,v_t$ of $D_0$. No arc enters a prefix $P$ from the remaining vertices. Summing degree imbalances over $P$ therefore yields
\begin{equation}\label{eq:prefix}
 |A_{D_0}(P,V(D_0)\setminus P)|=\sum_{v\in P}b_{D_0}(v)
\le \sum_{v\in P}\max\{0,b_{D_0}(v)\}
\le\iota(D_0)\le\iota(D) +2r.
\end{equation}
Use the following bags for the path decomposition. 
\[
 B_i=\{v_i\}\cup
 \{v_j:j<i\text{ and }v_jv_\ell\in E(\UG(D_0))\text{ for some }\ell\ge i\}.
\]
For the prefix \(P_{i-1}=\{v_1,\ldots,v_{i-1}\}\), every vertex in \(B_i\setminus\{v_i\}\) is the tail of an arc from \(P_{i-1}\) to the rest of the vertices. Choosing one such arc for each vertex gives distinct arcs, so Equation~\eqref{eq:prefix} implies \(|B_i|\le\iota(D) +2r+1\).
These bags cover all vertices and edges. A vertex \(v_j\) appears in consecutive bags, starting at \(B_j\) and ending at the bag indexed by its last later neighbor, or at \(B_j\) if it has no later neighbor. Add all $r$ vertices of $Z$ to every bag. This covers every remaining edge and gives width at most $\iota(D) +3r$. If $D_0$ is empty, the single bag $Z$ suffices.
\end{proof}
The imbalance bound controls the number of arcs
crossing each topological prefix; a bound on the
feedback vertex set alone would not provide this
control.

\begin{lemma}[$\star$]\label{lem:pathdp}
Given a width-$w$ path decomposition of $\UG(D)$,
$\EqSel(D,Q)$ can be decided in
$2^{O(w+h)}n^{O(1)}$ time, where $h=|Q|$.
If a solution exists, the algorithm constructs one.
\end{lemma}

The algorithm expands each bag by the in-neighbors
of its vertices and all equality endpoints, giving
bags of size at most $3(w+1)+2h$.
It then records the bag coloring and the colors
used so far, the latter ensuring nonconstancy.
The interval verification, table transitions,
and reconstruction appear in
Appendix~\ref{app:pathdp}.

\begin{theorem}\label{thm:eqsel-fpt}
On indegree-two digraphs, $\EqSel(D,Q)$ is FPT parameterized by $h+\iota(D)$. The algorithm constructs a nonconstant feasible coloring whenever one exists.
\end{theorem}
\begin{proof}
Apply Lemma~\ref{lem:constructive-packing}
with $t=h+2$.
If it returns cycles, complete them to a selector
$F$ as in Lemma~\ref{lem:many} and color one
component of $\UG(F)+Q$ one and the others zero.
Otherwise, it returns a feedback set of size
at most $R(h+2)$.
Lemma~\ref{lem:imbalance} gives a path decomposition
of width $w\le\iota(D)+3R(h+2)$.
Lemma~\ref{lem:pathdp} then decides feasibility
and constructs a solution in
$2^{O(\iota(D)+R(h+2)+h)}n^{O(1)}$ time.
Together with the computation in
Lemma~\ref{lem:constructive-packing},
this is FPT in $h+\iota(D)$.
\end{proof}

It remains to show that matching conflicts allow
$s_4(G)$ to bound both parameters of the selector
algorithm. 
\begin{proof}[Proof of Theorem~\ref{thm:s4}]
After singleton preprocessing, apply
Lemma~\ref{lem:branch}.
Matching conflicts give $2q\le m$, and the
chosen centered pairs use $2n$ distinct host
edges. Thus $D$ is an orientation of a spanning
subgraph of $G$, and
\begin{equation}\label{eq:surpluscount}
 4\eta=4q-4n\le2m-4n
 =\sum_v(d_G(v)-4)\le s_4(G).
\end{equation}
Each of the at most $2^\eta$ branches therefore
has $h\le\eta\le s_4(G)/4$.
Moreover, $d_D^+(v)\le d_G(v)-2$ gives
$\iota(D)\le s_4(G)$.
Theorem~\ref{thm:eqsel-fpt} solves every branch
constructively in FPT time in $s_4(G)$, and
Lemma~\ref{lem:branch} provides the answer and
witness to the original instance.
For $\Delta(G)\le5$, $s_4(G)$ is the number
of degree-five vertices.
\end{proof}

The matching assumption supplies the bounds
$4\eta\le s_4(G)$ and $\iota(D)\le s_4(G)$;
the auxiliary $\EqSel$ algorithm itself requires
no matching assumption.
Theorem~\ref{thm:overlap} extends the argument by
also accounting for edges that participate in
several conflicts.

The two surplus algorithms provide complementary
guarantees: the XP algorithm permits arbitrary
conflicts, whereas the FPT algorithm assumes
matching conflicts.
Even for matching instances surviving the singleton
test, bounded $\eta$ does not imply bounded $s_4(G)$.
Thus the FPT algorithm for $s_4(G)$ does not establish
fixed-parameter tractability in $\eta$.

\subsection{Overlapping conflicts and planar hosts}

We extend the preceding algorithm in two directions:
additional conflict incidences can be included in
the parameter, while planarity permits a
single-exponential dependence on $s_4(G)$.

We measure these additional conflict occurrences by
$
 p(\widehat G)=\sum_{e\in E(G)}\max\{0,d_{\widehat G}(e)-1\}.
$
Thus $p(\widehat G)=0$ exactly when $\widehat G$ is a matching, with isolated vertices permitted.

\begin{theorem}[$\star$]\label{thm:overlap}
For arbitrary conflicts, without locality, \CF{} is constructively FPT parameterized by $s_4(G)+p(\widehat G)$.
\end{theorem}

The selector algorithm remains unchanged.
The inequality $2q\le m+p(\widehat G)$ bounds the
conflict surplus. Moreover, at most $p(\widehat G)$
host edges can be selected at both endpoints,
bounding the additional imbalance caused by
opposite arcs. These observations give
\[
 4\eta\le s_4(G)+2p(\widehat G),
 \qquad
 \iota(D)\le s_4(G)+2p(\widehat G).
\]
Theorem~\ref{thm:eqsel-fpt} therefore applies to
every equality branch with the required parameter
bound. Appendix~\ref{app:overlap} gives the full proof.

For a conflict graph consisting of isolated vertices, edges, and $t_3$ triangles, $p(\widehat G)=3t_3$. Theorem~\ref{thm:overlap} therefore gives constructive FPT in $s_4+t_3$. When the conflicts are local, Section~\ref{sec:degree-hardness} develops an independent majority algorithm  with running time
$2^{O(s_4+t_2)}n^{O(1)}$,
where $t_2$ counts edge components.
Together, these algorithms give FPT in
$s_4+\min\{t_2,t_3\}$
(Corollary~\ref{cor:component-fpt} (ii)).

Planarity strengthens the packing alternative,
giving a single-exponential parameter dependence on $s_4(G)$.

\begin{theorem}[$\star$]\label{thm:planar-s4}
\CF{} on planar hosts with matching conflicts is solvable constructively in $2^{O(s_4(G))}n^{O(1)}$ time, even when conflicts are nonlocal. On planar hosts of maximum degree at most five, this is $2^{O(n_5)}n^{O(1)}$ time, where $n_5$ counts degree-five vertices.
\end{theorem}

The improvement uses the following polynomial-time
construction of a cycle packing and a directed
feedback vertex set with a linear size relationship.
Its derivation appears in
Appendix~\ref{app:planar-fpt}.

\begin{lemma}[$\star$]\label{lem:planar-packing}
For a planar digraph $D$, a family $\mathcal P$ of vertex-disjoint directed cycles and a directed feedback vertex set $Z$ with $|Z|\le12|\mathcal P|$ can be computed in polynomial time.
\end{lemma}

After singleton preprocessing, the auxiliary
digraph $D$ has a planar underlying graph.
If $|\mathcal P|\ge\eta+2$, the cycles yield a
cut as in Lemma~\ref{lem:many}.
Otherwise, $|Z|\le12(\eta+1)$.
Since $\eta\le s_4(G)/4$ and $\iota(D)\le s_4(G)$,
Lemma~\ref{lem:imbalance} gives a path decomposition
of width $O(s_4(G)+1)$.

Expand the bags by predecessors and add all
endpoints of the unchosen conflicts, at most
$4\eta$ vertices. Every original conflict can
then be tested directly in the dynamic program,
with bag size still $O(s_4(G)+1)$.
The packing and feedback-set computations are
polynomial-time, and the dynamic program takes
$2^{O(s_4(G))}n^{O(1)}$ time.
Appendix~\ref{app:planar-fpt} gives the complete
proof and the precise bag-size bounds.

\section{Conflict components}
\label{sec:degree-hardness}
We study local conflicts through their components:
triangle components yield a majority algorithm,
while mixtures of component types give hardness.
We then extend the algorithm to accommodate
degree surplus and exceptional components.

\subsection{Triangle components}

Under locality, triangle components of the conflict
graph yield another tractable case at maximum
degree four. Their host edges impose either
monochromatic triangles or majority equations,
which together determine whether a CF-cut exists.

\begin{theorem}\label{thm:triangles}
Let $G$ have maximum degree at most four, and suppose the conflicts are local and every component of $\widehat G$ is $K_1$ or $K_3$. Then a CF-cut can be found, or its nonexistence determined, in $O(n+m)$ time. %
\end{theorem}

The triangles here are components of the conflict graph. Their corresponding edges in $G$ need not form triangles.

\begin{observation}\label{obs:triangle-types}
Three distinct pairwise-intersecting edges of a simple graph either form a triangle or have a common endpoint.
\end{observation}
\begin{proof}
Write two of the edges as $va$ and $vb$. A third edge that intersects both either contains $v$ or is the edge $ab$.
\end{proof}

If the edges represented by a conflict triangle form a triangle in $G$, every feasible coloring makes its three vertices monochromatic: any coloring that is nonconstant on its three vertices cuts two of its edges. We call such a triangle a \emph{rigid triangle}. If the edges form a star with center $v$ and leaves $a,b,c$, at most one star edge may cross. Writing $\operatorname{Maj}$ for Boolean majority, this condition is
\begin{equation}\label{eq:majority}
 f(v)=\operatorname{Maj}(f(a),f(b),f(c)).
\end{equation}

\begin{figure}[tb]
\centering
\begin{tikzpicture}
\node[vertex] (a) at (0,0) {$a$};
\node[vertex] (b) at (2,0) {$b$};
\node[vertex] (c) at (1,1.7) {$c$};
\draw[draw=ink,thick] (a)--(b)--(c)--(a);
\node[align=center] at (1,-.85) {rigid host triangle\\$f(a)=f(b)=f(c)$};
\begin{scope}[xshift=7.2cm]
\node[vertex] (v) at (0,.55) {$v$};
\node[vertex] (a) at (-1.5,1.7) {$a$};
\node[vertex] (b) at (1.5,1.7) {$b$};
\node[vertex] (c) at (0,-.25) {$c$};
\draw[arr] (a)--(v);
\draw[arr] (b)--(v);
\draw[arr] (c)--(v);
\node[align=center] at (0,-1.05) {three-edge star\\$f(v)=\operatorname{Maj}(f(a),f(b),f(c))$};
\end{scope}
\end{tikzpicture}
\caption{The two configurations in $G$ represented by a local triangle component of $\widehat G$. The three displayed edges conflict pairwise. A triangle forces equality of the colors of its vertices, while a star requires its center to have the majority color of its leaves. The arrows on the star edges indicate the orientation
from majority inputs to their output, used in the proof;
the host graph itself is undirected.}
\label{fig:majority}
\end{figure}
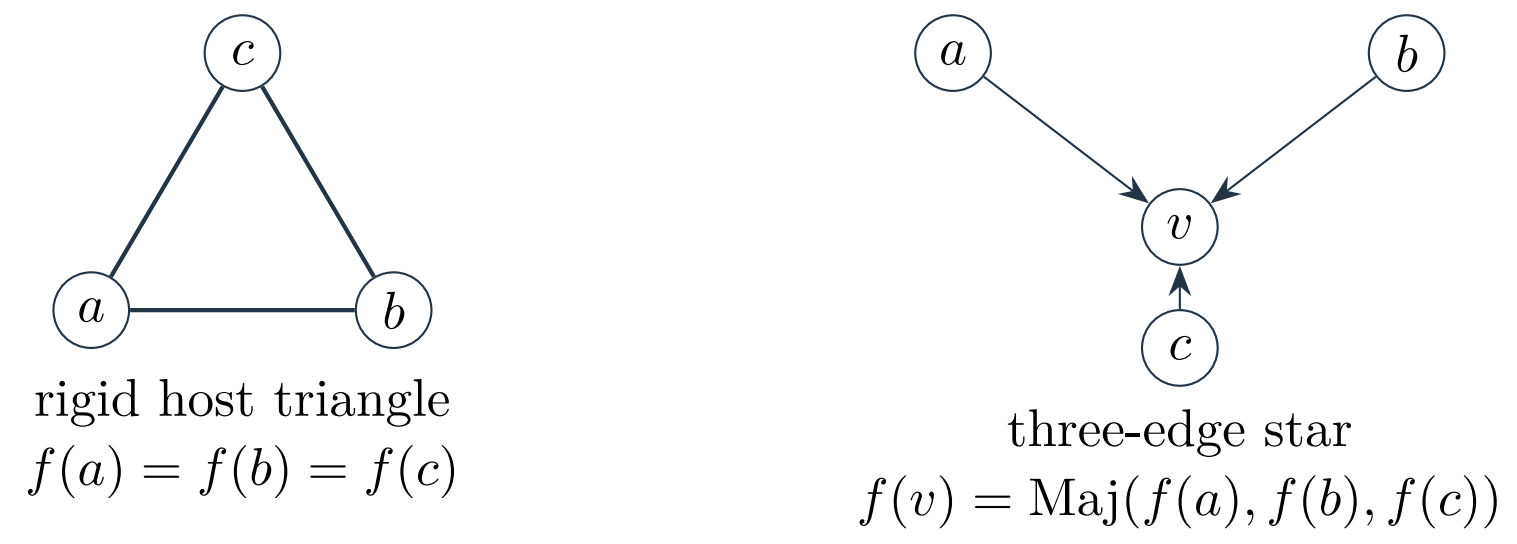

The following elementary propagation rule will be used here and in Section~\ref{sec:majority-extension}.

\begin{lemma}[$\star$]\label{lem:majority-propagation}
Suppose some vertices have fixed Boolean colors and every remaining vertex is the output of one majority equation with three specified inputs. There is a coloring satisfying all equations at the remaining vertices and extending the fixed colors. This coloring can be found in $O(n)$ time by starting all remaining vertices at zero and repeatedly changing an output to one when at least two of its inputs are one.
\end{lemma}

The proof and linear-time implementation are given
in Appendix~\ref{app:majority-propagation}.
The lemma guarantees only the equations whose
outputs are not fixed; equations at fixed vertices,
if present, must be checked separately.

\noindent We are now ready to prove the main theorem of this section.

\begin{proof}[Proof of Theorem \ref{thm:triangles}]
Let $T$ consist of the vertices belonging to rigid
triangles, and let $H$ be the graph on $T$ whose
edges are precisely those belonging to these
triangles. The connected components of $H$ are
called \emph{rigid classes}.
Every feasible coloring is constant on each rigid
class. Isolated vertices of $\widehat G$ impose
no constraints.

Test all single-vertex cuts and return any that
is conflict-free. Henceforth assume that all
such cuts fail. We show that a CF-cut exists
exactly when there are at least two rigid classes.

Conflict components have disjoint host-edge sets.
A rigid-triangle vertex cannot also center a
three-edge star, since this would require degree
at least five.
By singleton failure, every vertex outside $T$
centers a star, uniquely since two stars require
degree six.
Orient these stars towards their centers.
Each center has indegree three and outdegree
at most one, since its own star uses three
incident edges.
Finally, $T=\emptyset$ would give $n$
edge-disjoint three-edge stars, contradicting
$m\le2n$.

\begin{claim}\label{cl:triangle-criterion}
A nonconstant feasible coloring exists if and only if there are at least two rigid classes.
\end{claim}
\begin{claimproof}
If there are at least two classes, give one color one and give all the other classes color zero. Apply Lemma~\ref{lem:majority-propagation} to the star centers. Every rigid triangle is monochromatic and every star equation holds. The prescribed classes have both colors, so this gives a CF-cut.

Suppose instead that there is only one rigid class. By complementing colors, assume its color is zero. In a nonconstant feasible coloring, let $W$ be the nonempty set of one-colored vertices. Every vertex of $W$ is a star center and needs at least two inputs in $W$. Hence there are at least $2|W|$ oriented star edges inside $W$. 
Each vertex of $W$ has outdegree at most one, so there are at most $|W|$ such edges. This is a contradiction.
\end{claimproof}

Compute the components of $H$ and apply
Claim~\ref{cl:triangle-criterion} and
Lemma~\ref{lem:majority-propagation}.
Singleton cuts are tested by marking the common
endpoint of each conflict pair: precisely the
unmarked vertices give feasible cuts.
There are $O(m)$ conflicts, so this test,
component classification, computing the rigid
classes, and propagation all take $O(n+m)$ time.
\end{proof}

\subsection{Four-regular graphs with two-regular conflicts}

The preceding algorithm permits local triangle
components. Allowing four-cycle components as well
makes the problem NP-complete, even on exactly
$4$-regular hosts.

\begin{theorem}[$\star$]\label{thm:two-four}
\CF{} is NP-complete on simple connected $4$-regular
hosts with local exactly $2$-regular conflicts,
even when every component of $\widehat G$ is
a triangle or a four-cycle.
\end{theorem}

\subparagraph*{Proof overview.}
We reduce from nonconstant satisfiability of
$R$-constraints (Theorem~\ref{thm:SATR}).
Each variable occurrence receives a rigid triangle
whose edges form a conflict triangle.
Write $f(A)$ for the common color of triangle $A$.

An equality connector $z$ has two edges to
distinct vertices of each of two occurrence
triangles $A,B$.
Pair the edges to each triangle, then complete
these pairs to a conflict four-cycle.
The pairs force $f(A)=f(z)=f(B)$, making all
four edges monochromatic.
Linking occurrences cyclically enforces a
common variable color.

For $R(a,b,c)$, a clause connector $z$ has
one edge to each of the occurrence triangles
$A,C$ and two to $B$.
The two $B$-edges conflict, forcing $f(z)=f(B)$;
the edges to $A,C$ also conflict, enforcing
$R(f(A),f(B),f(C))$.
Complete these pairs to a conflict four-cycle.
Every added conflict contains a monochromatic
$B$-edge.

All connector conflicts are local.
Padding stars complete the host to degree four
without changing the possible assignments.
Appendix~\ref{app:two-four} gives the full
construction and correctness proof.

\subsection{A classification for local cluster conflict graphs}

Although conflict pairs and conflict triangles
separately permit polynomial-time algorithms,
allowing both gives NP-completeness.
We obtain a complete classification for local
cluster conflict graphs on hosts of maximum
degree four. Under these restrictions, every
clique component has order at most four.

\begin{theorem}[$\star$]\label{thm:component-dichotomy}
Fix any subset $\mathcal A\subseteq\{K_1,K_2,K_3,K_4\}$.
On simple connected graphs with $\Delta(G)\le4$
and local conflicts whose components belong to
$\mathcal A$, \CF{} is NP-complete if
$K_3\in\mathcal A$ and at least one of $K_2,K_4$
belongs to $\mathcal A$.
Hardness holds even on exactly $4$-regular hosts
with no isolated conflict vertices.
In all other cases, the problem is polynomial-time
solvable.
\end{theorem}

\subparagraph*{Proof overview.}
The polynomial cases follow from
Theorems~\ref{thm:triangles} and~\ref{thm:density}
and a singleton-cut counting argument.
For hardness, retain the host of
Theorem~\ref{thm:two-four}.
Replacing each conflict four-cycle by a suitably
chosen perfect matching preserves the equality
and clause constraints, giving components
$K_2,K_3$.
Alternatively, completing each four-cycle to
$K_4$ adds only local conflicts containing
a forced monochromatic edge, giving components
$K_3,K_4$.
Both changes preserve feasible colorings.
Appendix~\ref{app:component-dichotomy}
gives the full proof.

\subsection{An extension of the majority algorithm}\label{sec:majority-extension}
We extend the majority algorithm of
Theorem~\ref{thm:triangles} beyond degree four
and permit exceptional conflict components.
Combining this extension with the degree-surplus
algorithms of Theorems~\ref{thm:s4}
and~\ref{thm:overlap} then gives algorithms
parameterized by conflict-component counts.

Call a conflict component \emph{ordinary} if it
is $K_1$ or $K_3$, and \emph{exceptional} otherwise.
Let $B$ consist of the host vertices incident
with edges represented in exceptional components,
and let $b=|B|$.
Thus $b$ counts host vertices, and exceptional
components may have arbitrary structure.

\begin{theorem}[$\star$]\label{thm:exceptions}
For local conflicts on a simple connected graph
$G$, \CF{} can be solved constructively in
$2^{O(s_4(G)+b)}n^{O(1)}$ time.
After polynomial preprocessing, at most
$2^{2b+32s_4(G)}$ color assignments are tested,
each in polynomial time.
In particular, when $\Delta(G)\le4$, the bound
is $2^{O(b)}n^{O(1)}$.
\end{theorem}

\subparagraph*{Proof overview.}
After singleton preprocessing, mark as exceptional
every triangle component whose host edges meet
a vertex of degree greater than four.
The enlarged set $B'$ has size $O(b+s_4(G))$.
The unmarked triangle components yield rigid classes
and majority equations.
If $B'=\emptyset$, Theorem~\ref{thm:triangles} applies.

Suppose there are more rigid classes disjoint
from $B'$ than remaining majority centers in $B'$.
Give each such class a distinct label.
An unlabelled center outside $B'$ receives a label
when two of its inputs carry that label.
Each label defines a coloring: vertices carrying
it receive one, and all others receive zero.
Only majority equations at centers in $B'$ can
fail, and each center invalidates at most one
coloring. Hence some coloring gives a CF-cut.

Otherwise, there are only $O(b+s_4(G))$ rigid
classes. Enumerate consistent colors for these
classes and $B'$, considering only guesses using
both colors. Check the exceptional conflicts and
solve the remaining majority equations in polynomial
time along in-trees and directed cycles.
Return a cut if a guess extends, and report no
otherwise. Appendix~\ref{app:majority-extension}
gives the complete proof.

The following corollary collects the
component-count consequences.
Let $t_2,t_3,t_4$ count components isomorphic
to $K_2,K_3,C_4$, respectively. 
Table~\ref{tab:component-parameters} in
Appendix~\ref{app:majority-extension}
summarizes the following bounds.

\begin{corollary}\label{cor:component-fpt}
For local conflicts on simple connected hosts,
the following constructive bounds hold.
\begin{enumerate}[(i)]
\item If every conflict component belongs to
$\{K_1,K_2,K_3,C_4\}$, the running time is
$2^{O(s_4(G)+t_2+t_4)}n^{O(1)}$.

\item If every conflict component belongs to
$\{K_1,K_2,K_3\}$, the problem is FPT
parameterized by $s_4(G)+\min\{t_2,t_3\}$.
\end{enumerate}
\end{corollary}

\begin{proof}
A local $K_2$ component has three host endpoints,
and a local $C_4$ component has at most five.
Thus $b\le3t_2+5t_4$, giving~(i) by
Theorem~\ref{thm:exceptions}.
For~(ii), part~(i) gives FPT in $s_4(G)+t_2$,
while $p(\widehat G)=3t_3$ and
Theorem~\ref{thm:overlap} give FPT in $s_4(G)+t_3$.
Choose the algorithm corresponding to the
smaller component count.
\end{proof}

\section{Component-order connectivity and local conflicts}\label{sec:locality}

We now study locality under restrictions on the
components remaining after vertex deletion.

\subsection{An algorithm using a scope transversal}

The \emph{scope} of a local conflict $\{uv,vw\}$
is $\{u,v,w\}$.
A \emph{scope transversal} is a vertex set meeting
every scope; its minimum size is denoted by
$\tau_{\rm sc}$.

\begin{observation}\label{obs:coc-scope}
For local conflicts, every $2$-COC of $G$
is a scope transversal.
\end{observation}
\begin{proof}
Each scope induces a connected subgraph on three
vertices. If a $2$-COC missed it, these vertices
would remain in a component of order at least three.
\end{proof}

\begin{theorem}[$\star$]\label{thm:scope}
For local conflicts, \CF{} is solvable in
$O^*(2^k)$ time given a scope transversal of size
$k$, and in $O^*(3^{\tau_{\rm sc}})$ time when
one is not supplied.
In particular, a supplied $2$-COC of size $k$
gives the $O^*(2^k)$ bound; finding a minimum
$2$-COC and solving the problem takes
$O^*(3^{\coc_2(G)})$ time.
\end{theorem}

\subparagraph*{Proof overview.}
Enumerate the colors of the supplied scope
transversal $X$. Each conflict then involves
at most two unknown colors and can be expressed
as a 2-CNF formula.
Enforce nonconstancy by fixing each ordered pair
of distinct vertices to zero and one in turn.

To find a scope transversal, branch on the
three vertices of an uncovered scope.
For a $2$-COC, instead branch on three vertices
forming a connected subgraph in a component
of order at least three.
Increasing the budget until a set is found,
then applying the coloring algorithm, gives
the stated running times.
Appendix~\ref{app:scope} gives the full proof.

The algorithm allows an edge to participate in
arbitrarily many conflicts.
The next results show that locality and component
order two are essential to this guarantee,
even when every edge has exactly one
conflicting partner.

\subsection{Hardness for component-order deletion sets}

\begin{theorem}[$\star$]\label{thm:coc-hardness}
\CF{} is NP-complete on simple connected hosts
with exactly $1$-regular conflicts in each of
the following cases.
\begin{enumerate}[(i)]
\item All conflicts are local, and a set $X$
of six vertices is supplied such that $G-X$
consists of paths on three vertices, isolated
edges, and isolated vertices.

\item A $2$-COC $X$ of size six is supplied.
Every nonlocal conflict is between two edges
of $G-X$.
\end{enumerate}
\end{theorem}

The local construction also has bounded treewidth
and feedback-vertex number; Appendix~\ref{app:coc3}
gives the precise structural bounds.

\subparagraph*{Proof overview.}
Both reductions start from satisfiability of
$R$-constraints with constants
(Theorem~\ref{thm:SATR}).
Two rigid $K_5$ blocks represent zero and one.
Every outside vertex has a conflicting pair
of edges, one to each block; consequently,
every nonconstant feasible coloring gives
the blocks opposite colors.
Distinct vertices represent source occurrences.
Additional local conflicts fix constant occurrences
and make occurrences of each variable agree.

For part~(i), represent a constraint $R(a,b,c)$
by a path $a_Cb_Cc_C$ whose edges conflict,
imposing $R(f(a_C),f(b_C),f(c_C))$.
Deleting three designated vertices from each
constant block leaves the clause paths and
components of order at most two.
For part~(ii), split each middle occurrence into
two vertices forced to have the same color.
The clause becomes two disjoint edges forming
a nonlocal conflict that enforces the same relation.
Deleting the same six vertices now leaves
only isolated vertices and edges.
Both constructions have exactly $1$-regular
conflicts; full proofs appear in
Appendices~\ref{app:coc3} and~\ref{app:nonlocal2}.

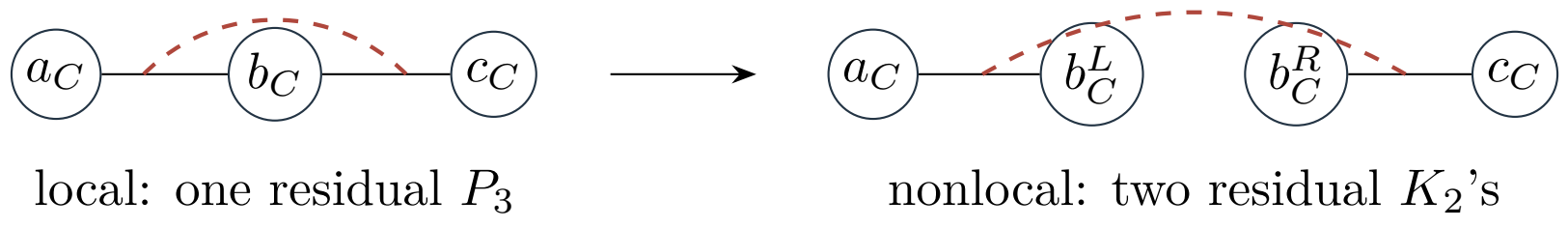
\begin{figure}[tb]
\centering
\begin{tikzpicture}
\foreach \n/\x/\l in {a/0/a_C,b/1.5/b_C,c/3/c_C}
 \node[vertex] (\n) at (\x,0) {$\l$};
\draw (a)--(b)--(c);
\draw[conflict] (.6,0) to[bend left=45] (2.4,0);
\node at (1.5,-.8) {local: one residual $P_3$};
\draw[-{Stealth}] (3.8,0)--(4.8,0);
\foreach \n/\x/\l in {a/5.6/a_C,b/7.1/b_C^L,d/8.5/b_C^R,c/10/c_C}
 \node[vertex] (\n) at (\x,0) {$\l$};
\draw (a)--(b);\draw (d)--(c);
\draw[conflict] (6.35,0) to[bend left=30] (9.25,0);
\node at (7.8,-.8) {nonlocal: two residual $K_2$'s};
\end{tikzpicture}
\caption{Clause components after deleting the
six-vertex set $X$.
The local construction leaves a $P_3$ whose
edges conflict. Splitting the middle occurrence
into two vertices forced to have the same color
leaves two $K_2$ components with a nonlocal
conflict. Dashed curves indicate conflicts.}
\label{fig:coc-boundary}
\end{figure}

Further host restrictions do not always yield
tractability. With local conflicts, \CF{} remains
NP-complete on cliques with matching conflicts
and on hosts with a universal vertex and exactly
$1$-regular conflicts.
Allowing several local conflicts per edge gives
hardness on interval graphs with a supplied
$3$-COC of size two.
In contrast, a supplied singleton $a$-COC permits
polynomial time for every fixed $a$, even without
locality.
Appendix~\ref{app:further-boundaries} summarizes
these results and their consequences for
structural parameters.

\section{Conclusions and open questions}\label{sec:conclusion}

We established complexity boundaries for CF-cut
through host degrees, conflict components, and locality.
Directed cycles connect the degree-four algorithm
to the surplus algorithms, while majority propagation
gives a second tractable structure for local triangle
components. Locality also yields FPT in $\coc_2(G)$,
although hardness persists for bounded $\coc_3(G)$.
Three questions arise naturally from these results.

\begin{enumerate}
\item Is CF-cut FPT in the conflict surplus $\eta$,
even for exactly $1$-regular conflicts?
After singleton preprocessing, this restriction
gives $m-2n=2\eta$.
Our XP algorithm leaves this open:
unlike $s_4(G)$, bounded $\eta$ does not bound
the auxiliary digraph's positive imbalance.

\item For matching conflicts on arbitrary hosts,
can a cut be constructed in
$2^{O(s_4(G))}n^{O(1)}$ time,
as for planar hosts?

\item What is the complexity on hosts of maximum
degree four with local conflicts whose components
are all four-cycles?
Allowing triangle components as well gives
NP-completeness.
\end{enumerate}

\subparagraph*{Note added.}
During final preparation, we learned of independent
work by Das et al.~\cite{DasEtAl2026} on CF-cut
and its size-bounded variant.
For arbitrary conflicts, they prove NP-hardness
at $\coc_2(G)\le2$ and polynomial-time solvability
at $\coc_2(G)\le1$; their dissociation number
is our $\coc_2$.
These results overlap with our study of component-order
 deletion parameters, but neither set of results subsumes
 the other.
Our hardness additionally imposes exactly
$1$-regular conflicts
(Theorem~\ref{thm:coc-hardness}(ii)),
while our $\coc_2$-FPT algorithm applies under
locality (Theorem~\ref{thm:scope}).

\subparagraph*{Use of generative AI.}
ChatGPT assistance was used to support this research. While the authors originated the ideas for the digraph equivalence, surplus algorithms, NP-completeness, and COC results, discussions with ChatGPT were used to seek specific reductions for the surplus algorithms and conflict component classification. ChatGPT assistance was also used to prepare the manuscript. The authors take responsibility for all mathematical
claims, proofs, references, and the final manuscript.

\label{page:mainend}
\clearpage
\label{page:references}

\clearpage
\appendix
\label{page:appendixstart}
\section{The conflict-surplus XP algorithm}\label{app:xp}
We prove Theorem~\ref{thm:eta} by giving an XP
algorithm for $\EqSel(D,Q)$ parameterized by
$h=|Q|$, then applying Lemma~\ref{lem:branch}.
The algorithm either detects sufficiently many
vertex-disjoint directed cycles or searches for
a small directed pattern recording the connections
between equality endpoints.
We first state the external algorithms, then
describe the patterns and our algorithm, prove
correctness, and analyze the running time.

\paragraph*{External algorithms.}
We use two algorithms of Reed, Robertson, Seymour,
and Thomas~\cite{RRST}.
For this purpose, a \emph{directed subdivision embedding}
of a digraph $J$ in $D$ injectively maps the vertices
of $J$ to $V(D)$ and replaces each non-loop arc by
a directed path between its endpoint images.
A loop is replaced by a directed cycle through
its vertex image. Distinct replacement paths or
cycles meet only at the images of their shared
endpoints, and their interiors avoid all other
vertex images.
Let $\psi:V'\to V(D)$ be an injective map,
where $V'\subseteq V(J)$.

\begin{theorem}[{Reed--Robertson--Seymour--Thomas,~\cite[(5.1)]{RRST}}]\label{ext:rrst-algorithms}
Fix $L\ge1$ and a digraph $J$, which may have loops. Given $D$ and $\psi: V' \rightarrow V(D)$ an injective map from $V' \subseteq V(J)$, there is a polynomial-time algorithm that returns one of three correct conclusions: the mapping $\psi$ extends to a directed subdivision embedding of $J$; it does not extend; or $D$ has $L$ vertex-disjoint directed cycles.
The polynomial exponents may depend on $L$ and $J$.
\end{theorem}

Taking $J$ to consist of $L$ vertices, each carrying
a loop, and setting $V'=\emptyset$ and $\psi$ to
the empty map gives the following consequence,
also stated in~\cite[(5.2)]{RRST}.

\begin{theorem}[{Reed--Robertson--Seymour--Thomas,~\cite[(5.2)]{RRST}}]\label{ext:rrst-corollary}
For every fixed integer $L\ge1$, there is a polynomial-time algorithm deciding whether an input digraph $D$ has $L$ vertex-disjoint directed cycles.
\end{theorem}

Once Theorem~\ref{ext:rrst-corollary} rules out
$L$ vertex-disjoint directed cycles, the third
conclusion of Theorem~\ref{ext:rrst-algorithms}
cannot arise. The latter therefore decides
whether the prescribed vertex map extends
to an embedding.

\subsection{The algorithm for
\texorpdfstring{$\EqSel(D,Q)$}{DCS(D,Q)}}
\label{sec:dscalg}
\paragraph*{Setup.}
We start with the digraph $D$ and equality set $Q$ as in Definition \ref{def:eqsel}. The digraph $D$ has two distinct in-neighbors at each vertex; opposite arcs are allowed. Its equality set is $Q$, with $h=|Q|$. We seek a selector $F$ such that $\UG(F) + Q$ is disconnected. 
By Lemma~\ref{lem:selector}, every weak component
of a selector consists of one directed cycle
with trees directed outwards from it.

For a selector $F$, let $H_F = \UG(F) + Q$ be the undirected graph obtained by forgetting the directions of its arcs and adding the equality edges $Q$. 
\begin{figure}[tb]
\centering
\begin{tikzpicture}
\node[vertex] (x) at (0,1) {$p_v$};
\node[vertex] (v) at (1,0) {$v$};
\node[vertex] (y) at (2,1) {$q_v$};
\draw[arr] (x)--(v);\draw[arr] (y)--(v);
\draw[conflict] (.45,.55) to[bend left=20] (1.55,.55);
\node[align=center] at (1,-.8)
  {$R(f(p_v),f(v),f(q_v))$};
\foreach \j/\xx in {1/4,2/6.6,3/9.2}{
 \node[vertex] (a\j) at (\xx,0) {};
 \node[vertex] (b\j) at (\xx+.9,0) {};
 \node[vertex] (c\j) at (\xx+.45,.85) {};
 \draw[arr] (a\j)--(b\j);\draw[arr] (b\j)--(c\j);\draw[arr] (c\j)--(a\j);
 \node[vertex] (t\j) at (\xx+.9,1.6) {};
 \draw[arr] (c\j)--(t\j);
}
\draw[conflict] (t1)--(t2) node[midway,above] {$Q$};
\node at (7.1,-.8) {one cycle in each selector component};
\end{tikzpicture}
\caption{A local conflict centered at $v$ is represented by two arcs into $v$. Selecting one incoming arc at each vertex gives components with one directed cycle each. Equality edges are dashed. In the right-hand illustration only the selected arcs are shown.}
\label{fig:selectors}
\end{figure}
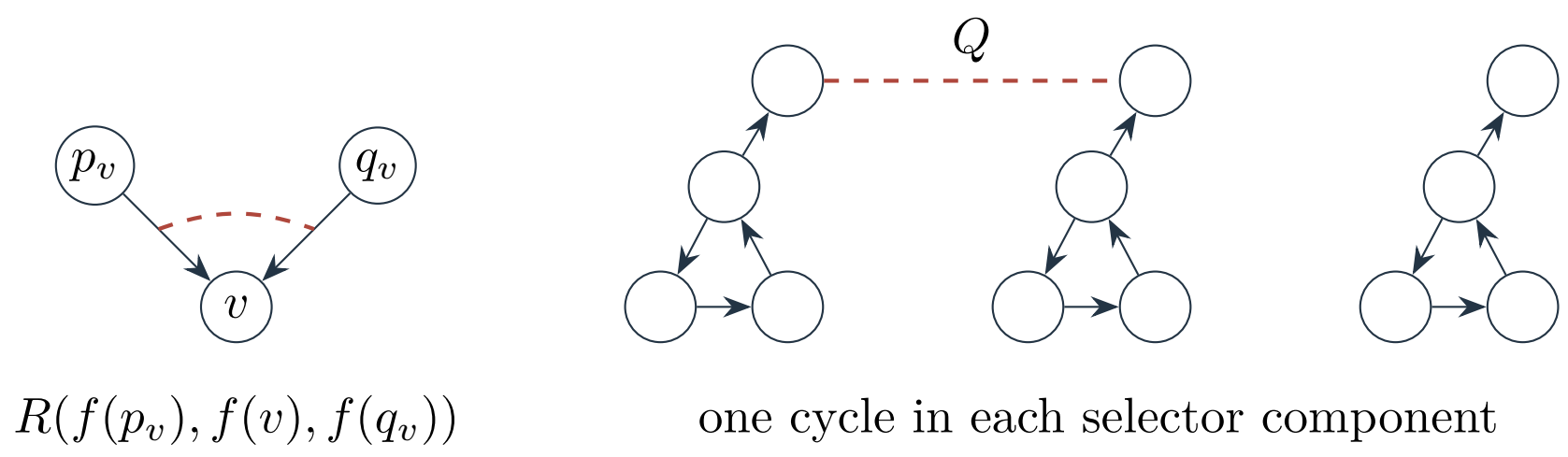
Let
\[
 T=V(Q),\qquad |T|\le2h,\qquad K=7h+1.
\]
We refer to the vertices of $T$ as the \emph{terminals}. We now define patterns; in an abstract sense these are the ways in which a selector $F$ may connect the terminals.

\begin{definition}[Candidate Pattern]\label{xp:def:pattern}
A \emph{candidate pattern} is a pair $(J,\psi)$, where $J$ is a
directed graph with loops permitted, satisfying the following
conditions:
\begin{enumerate}
    \item \textbf{Vertices and labels.}
    There is a designated set $T_J\subseteq V(J)$ of labelled
    vertices, and $\psi:T_J\to T$ is a bijection. Thus every
    terminal labels exactly one pattern vertex. We require
    \[
        \max\{1,|T|\}\le |V(J)|\le K.
    \]
    The vertices in $V(J)\setminus T_J$ are auxiliary vertices.

    \item \textbf{Arcs.}
    Every vertex of $J$ has exactly one incoming arc, with a
    loop contributing one to the indegree.

    \item \textbf{Disconnectedness after adding equalities.}
    Let $H_J$ be the undirected graph obtained from $\UG(J)$
    by adding the edge $\psi^{-1}(x)\psi^{-1}(y)$ for every
    equality pair $\{x,y\}\in Q$. We require $H_J$ to be
    disconnected.
\end{enumerate}
Loops have no effect on undirected connectivity. Each weak
component of $J$ contains exactly one directed cycle.
\end{definition}

We spell out the embedding conditions for a candidate
pattern $(J,\psi)$, including the requirement that
every terminal has its prescribed image.

\paragraph*{The embedding $\phi$.}
An \emph{embedding of $J$ in $D$ fixing the terminals} consists
of an injective map $\phi:V(J)\to V(D)$ and arc replacements
satisfying:
\begin{itemize}
    \item \textbf{Terminals.}
    $\phi(x)=\psi(x)$ for every $x\in T_J$.
Thus each labelled pattern vertex is mapped to the vertex
of $D$ specified by its label.

    \item \textbf{Arcs.}
    Each non-loop arc $u\to v$ is replaced by a directed path
    from $\phi(u)$ to $\phi(v)$; a loop at $v$ is replaced
    by a directed cycle through $\phi(v)$.

    \item \textbf{Intersections.}
    Each loop/arc replacement meets $\phi(V(J))$ only at the images
    of its pattern endpoints. Distinct replacements intersect
    only at the images of shared pattern endpoints.
\end{itemize}
The embedding need not be induced.

The algorithm first tests the sufficient condition
of Lemma~\ref{lem:many}. If that test fails, it
enumerates candidate patterns and tests whether
they admit embeddings fixing the terminals.

\paragraph*{$\EqSel(D,Q)$ Algorithm.}
\begin{enumerate}
 \item Test whether $D$ has $h+2$ vertex-disjoint directed cycles using Theorem \ref{ext:rrst-corollary}. If so, return \YES.
 \item Otherwise, enumerate the candidate patterns \((J,\psi)\) described above.
 \item For each candidate pattern \((J,\psi)\), apply Theorem~\ref{ext:rrst-algorithms} with \(L=h+2\) to determine whether \(\psi\) extends to a directed subdivision embedding of \(J\) in \(D\).
 Return \YES{} if any test succeeds, and \NO{} if all fail.
\end{enumerate}

The equality pairs are added as undirected edges
when forming $H_J$ to test disconnectedness;
they are not added as arcs to the pattern passed
to Theorem~\ref{ext:rrst-algorithms}.
In Step~3, we apply Theorem~\ref{ext:rrst-algorithms}
to the candidate pattern $(J,\psi)$ with $L=h+2$.
Since Step~1 has ruled out $h+2$ vertex-disjoint
directed cycles, the third conclusion cannot arise.
The theorem therefore decides whether $\psi$
extends to an embedding of $J$ in $D$.

\subsection{Why the algorithm is correct}
Step~1 is sound: preserve the incoming arcs of the $h+2$ cycles and choose incoming arcs arbitrarily elsewhere. Distinct preserved cycles belong to distinct selector components. Adding $h$ equality edges merges at most $h$ components, so at least two remain.

For the remaining steps, we prove the following characterization.
\begin{lemma}\label{xp:lem:pattern}
Suppose \(D\) has no \(h+2\) vertex-disjoint directed cycles. Then \(\EqSel(D,Q)\) has a solution if and only if some candidate pattern \((J,\psi)\) admits an embedding in \(D\) extending \(\psi\).
\end{lemma}
\begin{proof}
\paragraph*{Forward direction.}
Suppose \(\EqSel(D,Q)\) has a solution. Let $F$ be a selector for which $H_F  = \UG(F) + Q$ is disconnected. 
Since every weak component of \(F\) contains a directed cycle and \(D\) has no \(h+2\) vertex-disjoint directed cycles, \(F\) has at most \(h+1\) weak components. We demonstrate a way to obtain a candidate pattern $(J, \psi)$ that admits an embedding into $D$. 

Keep every cycle of $F$. Also keep, for each off-cycle terminal, the unique directed path from its component's cycle to that terminal. Delete the other tree vertices and arcs. Every off-cycle leaf of the remaining subdigraph is a terminal. Retain the following vertices:
\begin{enumerate}
 \item all $|T|$ terminals;
 \item every off-cycle vertex with at least two outgoing arcs \emph{after pruning};
 \item every cycle vertex from which a retained off-cycle tree departs;
 \item one arbitrary cycle vertex in each component containing no terminal.
\end{enumerate}
Suppress every other vertex, replacing its incoming and outgoing arcs by a single arc. A cycle with only one retained position becomes a loop. Every component retains a cycle position: it either has a terminal on its cycle, has a departure leading to an off-cycle terminal, or receives the representative in item~4. This is illustrated in Figure \ref{fig:xp-pruning-compact}.

There are at most $|T|$ off-cycle branch vertices. Consider the components with a nonempty retained off-cycle tree, and contract their cycles to roots. Every leaf of each resulting outward tree is a terminal. The number of branch vertices in a rooted tree is at most its number of leaves minus one. Components with no off-cycle tree contribute no off-cycle branch vertices.

There are also at most $|T|$ cycle departures: choose a descendant terminal for each departure; trees attached at distinct cycle positions are disjoint, so these chosen terminals are distinct. The representatives in item~4 number at most $h+1$, the upper bound on the number of weak components. Thus the resulting pattern satisfies
\[
 |V(J)|\le |T|+|T|+|T|+(h+1) \le6h+(h+1)=7h+1.
\]
The categories may overlap, which only decreases the size of their union.

The pattern still has indegree one everywhere. Give each terminal its original label and give the other retained vertices auxiliary names. Undoing the suppressions exhibits an embedding in $D$ fixing the terminals. Pruning and suppression preserve every selector component and the component containing each terminal. Therefore adding $Q$ connects the pattern components in exactly the same way as it connected the selector components. Since $H_F$ is disconnected, so is $H_J$, making $J$ a candidate.

\begin{figure}[tb]
    \centering
    \begin{tikzpicture}[
  x=1cm,y=1cm,font=\small,
  cfarc/.style={-{Stealth[length=1.35mm,width=1mm]},line width=.5pt},
  cfeq/.style={densely dashed,line width=.65pt},
  cfdiscard/.style={-{Stealth[length=1.1mm,width=.8mm]},draw=black!35,line width=.45pt},
  cfterm/.style={circle,draw,fill=white,double,double distance=.45pt,
    minimum size=4.8mm,inner sep=.2pt},
  cfbranch/.style={rectangle,draw,fill=white,minimum size=4.3mm,inner sep=.2pt},
  cfdepart/.style={diamond,aspect=1.15,draw,fill=white,minimum size=5.3mm,inner sep=.1pt},
  cfrep/.style={regular polygon,regular polygon sides=6,draw,fill=white,
    minimum size=5mm,inner sep=.2pt},
  cfdot/.style={circle,fill=black,minimum size=1.5mm,inner sep=0pt},
  cffaded/.style={circle,draw=black!35,fill=white,minimum size=1.5mm,inner sep=0pt}
]
\path[use as bounding box] (-.3,-1.05) rectangle (13.6,4.05);

\node[anchor=west] at (0,3.88) {\textbf{(a)} Selector $F$};
\node[anchor=west] at (7.05,3.88) {\textbf{(b)} Pattern $(J,\psi)$};

\node[cfdepart] (fv1) at (.65,.7) {$v_1$};
\node[cfdot] (fc1) at (1.75,.12) {};
\node[cfdepart] (fv2) at (2.85,.7) {$v_2$};
\node[cfdot] (fc2) at (1.75,1.16) {};
\draw[cfarc] (fv1)--(fc1);
\draw[cfarc] (fc1)--(fv2);
\draw[cfarc] (fv2)--(fc2);
\draw[cfarc] (fc2)--(fv1);

\node[cfdot] (fs1) at (.65,1.5) {};
\node[cfbranch] (fw) at (.65,2.18) {$w$};
\node[cfterm] (fx1) at (.02,3.02) {$x_1$};
\node[cfdot] (fs2) at (1.3,2.6) {};
\node[cfterm] (fx2) at (1.94,3.02) {$x_2$};
\node[cfdot] (fs3) at (2.85,1.93) {};
\node[cfterm] (fx3) at (2.85,3.02) {$x_3$};
\draw[cfarc] (fv1)--(fs1);
\draw[cfarc] (fs1)--(fw);
\draw[cfarc] (fw)--(fx1);
\draw[cfarc] (fw)--(fs2);
\draw[cfarc] (fs2)--(fx2);
\draw[cfarc] (fv2)--(fs3);
\draw[cfarc] (fs3)--(fx3);
\node[cffaded] (fp1) at (.08,1.84) {};
\node[cffaded] (fp2) at (-.08,2.3) {};
\draw[cfdiscard] (fs1)--(fp1);
\draw[cfdiscard] (fp1)--(fp2);

\node[cfterm] (fx4) at (4.3,1.46) {$x_4$};
\node[cfdot] (fc3) at (3.86,.38) {};
\node[cfdot] (fc4) at (4.74,.38) {};
\draw[cfarc] (fx4)--(fc3);
\draw[cfarc] (fc3)--(fc4);
\draw[cfarc] (fc4)--(fx4);

\node[cfrep] (fz) at (5.93,1.46) {$z$};
\node[cfdot] (fc5) at (5.49,.38) {};
\node[cfdot] (fc6) at (6.37,.38) {};
\draw[cfarc] (fz)--(fc5);
\draw[cfarc] (fc5)--(fc6);
\draw[cfarc] (fc6)--(fz);
\node[cffaded] (fp3) at (5.93,2.17) {};
\node[cffaded] (fp4) at (5.42,2.83) {};
\node[cffaded] (fp5) at (6.4,2.83) {};
\draw[cfdiscard] (fz)--(fp3);
\draw[cfdiscard] (fp3)--(fp4);
\draw[cfdiscard] (fp3)--(fp5);

\draw[cfeq] (fx1.north) .. controls (.38,3.55) and (1.58,3.55) .. (fx2.north);
\draw[cfeq] (fx3.east) to[out=0,in=90] (fx4.north);
\node at (1.75,-.17) {$F_1$};
\node at (4.3,-.17) {$F_2$};
\node at (5.93,-.17) {$F_3$};

\begin{scope}[xshift=7.05cm]
\node[cfdepart] (jd1) at (.65,.72) {$d_1$};
\node[cfdepart] (jd2) at (2.85,.72) {$d_2$};
\node[cfbranch] (jb) at (.65,2.15) {$b$};
\node[cfterm] (ja1) at (.02,3.02) {$a_1$};
\node[cfterm] (ja2) at (1.94,3.02) {$a_2$};
\node[cfterm] (ja3) at (2.85,3.02) {$a_3$};
\draw[cfarc] (jd1) to[bend right=20] (jd2);
\draw[cfarc] (jd2) to[bend right=20] (jd1);
\draw[cfarc] (jd1)--(jb);
\draw[cfarc] (jb)--(ja1);
\draw[cfarc] (jb)--(ja2);
\draw[cfarc] (jd2)--(ja3);
\node[cfterm] (ja4) at (4.3,1.46) {$a_4$};
\node[cfrep] (jr) at (5.93,1.46) {$r$};
\draw[cfarc] (ja4) edge[loop below,min distance=8mm] (ja4);
\draw[cfarc] (jr) edge[loop below,min distance=8mm] (jr);
\draw[cfeq] (ja1.north) .. controls (.38,3.55) and (1.58,3.55) .. (ja2.north);
\draw[cfeq] (ja3.east) to[out=0,in=90] (ja4.north);
\node at (1.75,-.17) {$J_1$};
\node at (4.3,-.17) {$J_2$};
\node at (5.93,-.17) {$J_3$};
\end{scope}

\node[cfterm,minimum size=3.5mm] at (.08,-.66) {};
\node[anchor=west] at (.33,-.66) {terminal};
\node[cfbranch,minimum size=3.5mm] at (2.58,-.66) {};
\node[anchor=west] at (2.85,-.66) {off-cycle branch};
\node[cfdepart,minimum size=4mm] at (6.33,-.66) {};
\node[anchor=west] at (6.64,-.66) {cycle departure};
\node[cfrep,minimum size=3.8mm] at (10.1,-.66) {};
\node[anchor=west] at (10.38,-.66) {cycle representative};
\end{tikzpicture}
    \caption{Pruning and suppression for $h=2$ and
    $Q=\{\{x_1,x_2\},\{x_3,x_4\}\}$.
    (a) A selector $F$, with pale branches indicating the parts
    removed by pruning.
    (b) Suppressing the remaining small filled vertices gives
    the pattern $J$, with terminal labels $\psi(a_i)=x_i$.
    Dashed lines depict equalities, not directed arcs.
    Cycles with a single retained vertex become loops.
    The eight retained vertices comprise four terminals,
    one off-cycle branch, two cycle departures, and one
    cycle representative.}
    \label{fig:xp-pruning-compact}
\end{figure}
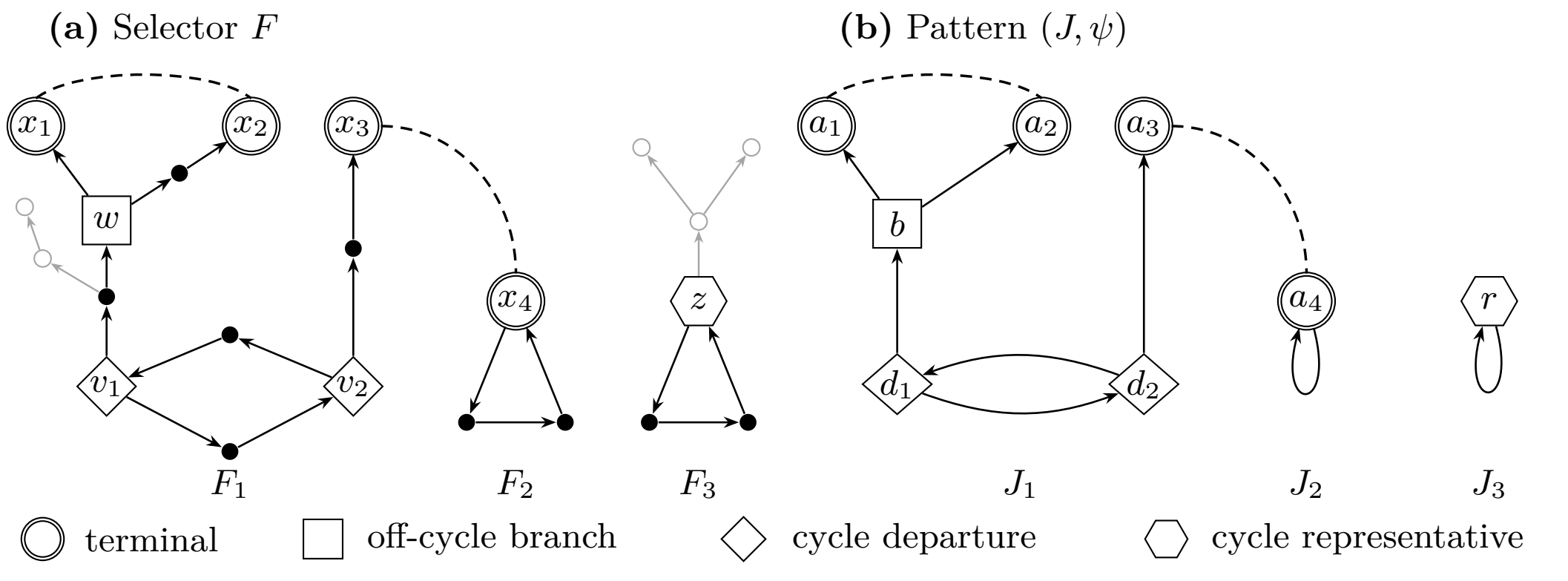

\begin{figure}[tb]
    \centering
    \begin{tikzpicture}[
  x=1cm,y=1cm,font=\small,
  cfcomp/.style={circle,draw,fill=white,minimum size=6.5mm,inner sep=0pt},
  cfeq/.style={densely dashed,line width=.65pt},
  cfregion/.style={rounded corners=5pt,draw=black!40,line width=.4pt}
]
\path[use as bounding box] (-.3,-.8) rectangle (11.05,1.62);
\draw[cfregion] (-.2,-.42) rectangle (6.7,1.43);
\draw[cfregion] (8.5,-.42) rectangle (10.6,1.43);
\node[cfcomp] (qj1) at (1.0,.1) {$J_1$};
\node[cfcomp] (qj2) at (5.45,.1) {$J_2$};
\node[cfcomp] (qj3) at (9.55,.1) {$J_3$};
\draw[cfeq] (qj1) edge[loop above,min distance=5.5mm]
  node[above,inner sep=1pt] {$\{a_1,a_2\}$} (qj1);
\draw[cfeq] (qj1)--node[above,inner sep=2pt] {$\{a_3,a_4\}$} (qj2);
\node at (3.25,-.67) {one component of $H_J$};
\node at (9.55,-.67) {another component};
\end{tikzpicture}
    \caption{Connectivity after adding the equalities in
    Figure~\ref{fig:xp-pruning-compact}. Each node represents
    an entire weak component of $J$. The pair $\{a_1,a_2\}$
    is internal to $J_1$, while $\{a_3,a_4\}$ joins $J_1$
    and $J_2$. Component $J_3$ remains separate, so $H_J$
    is disconnected. The dashed loop denotes an internal
    equality, not a directed cycle.}
    \label{fig:xp-equalities-compact}
\end{figure}
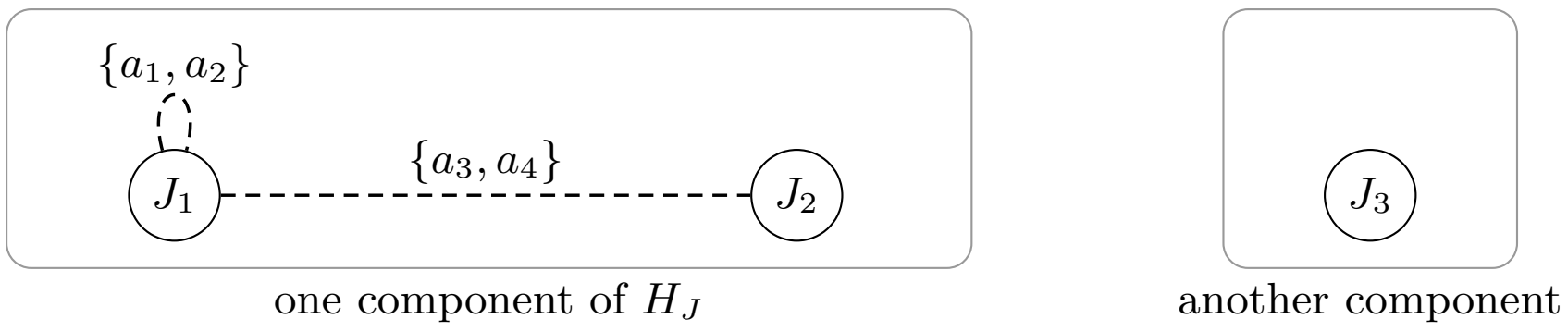

\paragraph*{Reverse direction.}
Suppose there is a candidate pattern $(J, \psi)$ that admits an embedding in $D$. Let $R$ be the subdigraph of $D$ occupied by the embedding. Every vertex of $R$ has exactly one incoming arc within $R$, including the internal vertices of the paths and cycles replacing pattern arcs. Choose these incoming arcs, and choose an arbitrary incoming arc at every vertex outside $R$, obtaining a spanning selector $F$.

Two embedded pattern components cannot merge in $F$: each retains a directed cycle, whereas a selector component contains only one cycle by Lemma \ref{lem:selector}. Additional selector components may arise outside $R$. Every equality endpoint in $T$ already lies in $R$, because every terminal was included and fixed. Consequently the equality edges join the components containing the embedded pattern exactly as they do in $H_J$, and no additional component can supply a further equality connection. Since $H_J$ is disconnected, $H_F$ is disconnected as well. 
\end{proof}

\subsection{Running time and the conflict-surplus bound}
The enumeration involves a number of patterns depending only on $h$. To see this directly, fix an order of the terminal labels. For each $s$ with $\max\{1,|T|\}\le s\le K$, use those $|T|$ labels together with $s-|T|$ fixed auxiliary names. Choosing the predecessor of each vertex gives all $s^s$ indegree-one patterns on this vertex set, including loops. Check the disconnectedness of $H_J$ for each. Hence the number of candidates considered is at most
\[
 \sum_{s=\max\{1,|T|\}}^{K}s^s\le K^{K+1} = (7h+1)^{7h+2}.
\]
There is no need to remove isomorphic duplicates or guess positions of terminals in $D$: their names already prescribe their images.

For fixed \(h\), the number \(h+2\) of disjoint directed cycles sought, the maximum pattern size \(7h+1\), and the number of calls to Theorems~\ref{ext:rrst-algorithms} and \ref{ext:rrst-corollary} all depend only on \(h\), not on the size of \(D\). By Theorems~\ref{ext:rrst-algorithms} and \ref{ext:rrst-corollary} each call runs in polynomial time for fixed \(h\). Since the number of calls also depends only on \(h\), the total running time is \(a(h)n^{c(h)}\) for computable functions \(a\) and \(c\).
This gives an XP algorithm for \(\EqSel(D,Q)\) parameterized by \(h\). When $h=0$, Step~1 tests for two disjoint cycles. If none exist, $K=1$ and no pattern has disconnected $H_J$, so the algorithm correctly returns \NO.

\paragraph*{Returning to conflict-free cuts.}
Let the host have $n$ vertices and let $q$ be the number of conflicts. First test all single-vertex cuts. If none works, each vertex centers a conflict between two incident edges. Choose one such pair at each vertex; these $n$ pairs are distinct because the host is simple. They give the predecessor digraph $D$ with two incoming arcs at each vertex. If an edge of $G$ is chosen at both endpoints, retain both opposite arcs.

There are $q-n=\eta$ remaining conflicts.
For each remaining pair $\{ab,cd\}$, branch on
requiring either $f(a)=f(b)$ or $f(c)=f(d)$. 
This is exactly the condition that the two conflicting edges do not both cross the cut. After duplicate equalities are removed, each of the at most $2^\eta$ branches has $h\le\eta$ equalities. A conflict-free cut exists exactly when one branch admits a nonconstant coloring satisfying its predecessor constraints and equalities.

For each of the at most \(2^\eta\) branches, apply the \(\EqSel(D,Q)\) algorithm of Section \ref{sec:dscalg} to the resulting instance \(\EqSel(D,Q)\). Accept the original instance if at least one branch is accepted; otherwise, reject it. 
Taking maxima of $a(h),c(h)$ over $h\le\eta$ and including the polynomial preprocessing gives time $a'(\eta)n^{c'(\eta)}$ for computable functions $a',c'$. This proves XP in $\eta=\max\{0,q-n\}$ for arbitrary conflicts. If $q<n$, a single-vertex cut must work; if $q=n$, the only remaining test is the test for two vertex-disjoint cycles using either Theorem \ref{ext:mccuaig} or Theorem \ref{ext:rrst-corollary}. The algorithm enumerates small abstract patterns and lets Theorem \ref{ext:rrst-algorithms} find their possible subdivisions; it never enumerates all directed paths or cycles of $D$.

\section{Details of the fixed-parameter algorithms}
\label{app:fpt}

This appendix supplies the proofs deferred from
Section~\ref{sec:surplus}.
We derive the constructive packing alternative
and describe the dynamic program on expanded bags.
We then give the overlap extension and the planar
refinement, and conclude with the tower-bounded
running-time observation for decision.

\subsection{The constructive packing alternative}
\label{app:fpt-packing}

We prove Lemma~\ref{lem:constructive-packing}.
We need an algorithm that either finds sufficiently many cycles or finds a bounded directed feedback vertex set, since the subsequent algorithm uses these
objects explicitly. Both outcomes must supply the corresponding objects. Write $\nu(D)$ for the maximum number of vertex-disjoint directed cycles. We collect the external results needed in one statement. A function is \emph{tower-bounded} here if it is bounded by a tower of exponentials of fixed height with a polynomial in the argument at the top.

\begin{theorem}[Directed cycle packing and feedback vertex sets]\label{ext:cycle-tools}
The following results hold for digraphs $D$.
\begin{enumerate}[(i)]
\item \textup{(Reed--Robertson--Seymour--Thomas~\cite[(1.3)]{RRST}; Hatzel--Kreutzer--Milani--Muzi~\cite[Theorem~1.1]{HKMM}.)}
There is a computable nondecreasing function $F$ such that, for every integer $k\ge1$, either $\nu(D)\ge k$, or $D$ has a directed feedback vertex set of size at most $F(k)$. The function $F$ may be chosen tower-bounded.
\item \textup{(Grohe--Gr\"uber~\cite[Definition~4 and Theorem~9]{GroheGrueber}.)}
There are a computable nondecreasing unbounded function $g$ and a polynomial-time algorithm on input $(D,k)$ which returns at least $g(k)$ vertex-disjoint directed cycles whenever $\nu(D)\ge k$. If $\nu(D)<k$, it may return a smaller valid packing or report failure.
\item \textup{(Chen--Liu--Lu--O'Sullivan--Razgon~\cite{ChenEtAl}.)}
Given $D$ and an integer $k\ge0$, one can find a directed feedback vertex set of size at most $k$, or correctly report that no such set exists, in time $4^k k!(|V(D)|+|A(D)|)^{O(1)}$.
\end{enumerate}
\end{theorem}

Part (i) is the cycle-packing versus feedback-vertex-set theorem of Reed et al. \cite{RRST}, with the improved bound of Hatzel et al. \cite{HKMM}. Part~(ii) restates the cited approximation guarantee: its ratio $\rho$ has $k/\rho(k)$ nondecreasing and unbounded, so take $g(k)=\lfloor k/\rho(k)\rfloor$. If the promise fails, checking the output and treating invalid output as failure gives the stated form. Combining the existence statement in~(i) with the two algorithms in parts~(ii)--(iii) proves Lemma~\ref{lem:constructive-packing}.

\begin{proof}[Proof of Lemma~\ref{lem:constructive-packing}]
Choose $K(t)=\min\{k:g(k)\ge t\}$ and run the algorithm of Theorem~\ref{ext:cycle-tools}(ii) with target $k= K(t)$. If it returns at least $t$ cycles, keep any $t$ of them. Otherwise its guarantee implies $\nu(D)<K(t)$, so Theorem~\ref{ext:cycle-tools}(i) guarantees the existence
of a directed feedback vertex set of size at most
$R(t)=F(K(t))$. Theorem~\ref{ext:cycle-tools}(iii), with budget $R(t)$, finds such a set. Computing the target and running that algorithm take a computable parameter-dependent factor times a polynomial in the input size.
\end{proof}

The arc-disjoint counterpart of this alternative already appears in~\cite[Lemma~2]{GJSW}; the preceding proof states the vertex-disjoint form and its external ingredients explicitly.

\subsection{Dynamic programming on expanded bags}
\label{app:pathdp}

We prove Lemma~\ref{lem:pathdp} using the coloring
equivalence of Lemma~\ref{lem:selector}.
We first expand the bags so that every constraint
is contained in a bag, then record the bag coloring
and the set of colors used so far.

\begin{proof}
We first modify the decomposition so that the vertices of each constraint occur together in some bag. For every bag $B$, replace it by
\[
 B^*=\bigcup_{v\in B}\bigl(\{v\}\cup N_D^-(v)\bigr),
\]
and add all endpoints of pairs in $Q$ to every bag. Each new bag has at most $3(w+1)+2h$ vertices. The predecessor constraint at $v$ is covered by every expanded bag that originally contained $v$, and every equality is covered as well.

We check that this is still a path decomposition. Since bags only gain vertices, vertex and edge coverage are preserved. It remains to check the interval property. Under the replacement \(B\mapsto B^*\), the bags containing a vertex \(x\) form the union of its original interval and the original intervals of its out-neighbors. Each of the latter intervals intersects the interval of $x$, because some original bag contains both endpoints of the corresponding edge. Their union is therefore an interval. Adding an equality endpoint to all bags also preserves this property.

\paragraph*{The table.}
We seek a nonconstant binary coloring on $V(D)$ that satisfies the conditions in Lemma \ref{lem:selector}.
We refine the expanded decomposition into a nice path
decomposition, retaining its bags without increasing its width
\cite[Lemma~7.2]{Cygan2015}. In such a decomposition, consecutive bags differ by the introduction or deletion of one vertex, and the bags at the beginning and the end are empty. 
Assign each constraint to a bag containing all its vertices. 
Each constraint is tested when its assigned bag is processed; the table for that bag records the states remaining after these tests.
For a bag $B_i$, a coloring $\sigma:B_i\to\bits$, and a set $U\subseteq\bits$, let $M_i(\sigma,U)$ be one if there is a coloring of the vertices introduced so far that agrees with $\sigma$, satisfies the constraints processed so far, and uses exactly the colors in $U$. Otherwise, let it be zero.

Initially only the entry for the empty coloring and $U=\emptyset$ is one. The entries are updated as follows.
\begin{enumerate}
\item When a vertex $x$ is introduced, extend each feasible coloring by either value $a\in\bits$ and replace $U$ by $U\cup\{a\}$.
\item When a vertex $x$ is forgotten, restrict the bag coloring to the remaining vertices and retain the same set $U$. An entry is one if at least one of its predecessor entries is one.
\item When a constraint assigned to the current bag is processed, set to zero every entry whose coloring violates that constraint.
\end{enumerate}
We accept if the final empty bag has an entry with $U=\bits$.

\paragraph*{Correctness and running time.}
The meaning of the table follows by induction along the decomposition. A vertex is introduced only once, by the interval property. Each constraint is tested while all its vertices are present. The set $U$ records colors that have occurred, including at forgotten vertices, so the final test is exactly the nonconstancy requirement.

There are at most $4\cdot2^{3(w+1)+2h}$ entries per bag. The refinement, constraint tests and table updates add only a polynomial factor. The running time is therefore $2^{O(w+h)}n^{O(1)}$. Storing a predecessor for every successful entry allows us to recover a coloring.
\end{proof}

\paragraph*{Comparing the surplus parameters $\eta$ and $s_4(G)$.}
The XP algorithm parameterized by $\eta$ and the FPT algorithm
parameterized by $s_4(G)$ provide complementary guarantees.
The former applies to arbitrary conflicts, whereas the latter
requires matching conflicts. For matching instances surviving
the singleton test, Equation~\eqref{eq:surpluscount} gives
$4\eta\le s_4(G)$. The reverse implication fails: degrees below
four can offset degrees above four in $2m-4n$. For example,
start with a regular tournament on five vertices, each having
indegree two and outdegree two. Add $k$ vertices with incoming
arcs from the same two original vertices, and pair the incoming
edges at every vertex. These matching instances have $q=n$,
$m=2n$, and $s_4(G)=2k$. Consequently, the FPT result parameterized by $s_4(G)$
does not by itself establish fixed-parameter tractability
with respect to $\eta$ or to nonnegative $m-2n$. When conflicts may
overlap, even bounded $s_4(G)$ need not imply bounded $\eta$.
The next subsection proves
Theorem~\ref{thm:overlap}, which accounts for
additional conflict incidences when bounding
both the conflict surplus and the imbalance
of the auxiliary digraph.

\subsection{Proof of Theorem~\ref{thm:overlap}}
\label{app:overlap}

The selector algorithm also applies when an edge
participates in several conflicts.
We show that the additional conflict incidences
bound both the extra equality choices and the
imbalance caused by edges selected at both endpoints.

\begin{proof}[Proof of Theorem~\ref{thm:overlap}]
Test single-vertex cuts first. Otherwise choose one centered pair per vertex of $G$, and let $\eta=q-n\ge0$.  We have \[
2q = \sum_{e\in E(G)} d_{\widehat G}(e)\le \sum_{e\in E(G)} \left ( 1 + \max\{0,d_{\widehat G}(e)-1\}\right ) = m + p(\widehat G).
\]
Therefore, 
\[
 4\eta = 4q - 4n \leq 2m +2p(\widehat G) -4n \le s_4(G)+2p(\widehat G),
\]
where the last inequality follows from $2m - 4n = \sum_{v \in V(G)} (d_G(v) - 4) \leq s_4(G)$.
Form the indegree-two digraph $D$, retaining opposite arcs when an edge is selected at both endpoints. Let $d$ count the doubly selected edges of $G$. Each such edge appears in two distinct centered conflicts, so has conflict degree at least two and contributes at least one to $p(\widehat G)$. Hence $d\le p(\widehat G)$.

For a vertex $v$, let $d_v$ count incident doubly selected edges. Its two incoming selected edges normally leave at most $d_G(v)-2$ possible outgoing edges, but doubly selected edges may also contribute outgoing arcs. Thus
\[
 d_D^+(v)\le d_G(v)-2+d_v,
 \qquad
 \iota(D)\le s_4(G)+\sum_v d_v
             =s_4(G)+2d\le s_4(G)+2p(\widehat G).
\]
Branching on the $\eta$ unchosen conflicts gives at most $\eta$ equalities per branch. Both the number of branches and the parameter of Theorem~\ref{thm:eqsel-fpt} are bounded by functions of $s_4+p(\widehat G)$. Its constructive algorithm and Lemma~\ref{lem:branch} complete the proof.
\end{proof}

\subsection{The planar refinement}
\label{app:planar-fpt}

We derive Lemma~\ref{lem:planar-packing} from
fractional cycle-packing bounds, then prove
Theorem~\ref{thm:planar-s4}.
The essential improvement is that the packing
and feedback vertex set are computed in polynomial
time and satisfy a linear size relationship.

Planarity allows us to find, in polynomial time, either
$\eta+2$ vertex-disjoint directed cycles or a directed feedback
vertex set of size at most $12(\eta+1)$, replacing the general
bound $R(\eta+2)$.
In the latter case, this linear bound yields a path decomposition
of width $O(s_4(G)+1)$ and hence the single-exponential running
time $2^{O(s_4(G))}n^{O(1)}$.
 We first state the two approximation bounds from which this follows. Let $\mathcal C(D)$ be the family of directed cycles of $D$, and define the fractional vertex cycle-packing value by
\[
 \lambda(D)=\max\left\{\sum_{C\in\mathcal C(D)}x_C:
       \sum_{\substack{C\in\mathcal C(D)\\v\in V(C)}}x_C\le1
       \text{ for every }v,\quad x_C\ge0\right\}.
\]
For an acyclic digraph this value is zero.

\begin{theorem}[Planar directed-cycle approximation bounds]\label{ext:planar-tools}
For a planar digraph $D$, the following sets can be computed in polynomial time.
\begin{enumerate}[(i)]
\item \textup{(Schlomberg--Thiele--Vygen~\cite[Theorem~5(a), Propositions~8--9]{STV}, directed-cycle case.)}
A family $\mathcal P$ of vertex-disjoint directed cycles with $|\mathcal P|\ge\lambda(D)/5$.
\item \textup{(Berman--Yaroslavtsev~\cite[Theorems~3.1 and~4.2]{BY}, unweighted directed-cycle case.)}
A directed feedback vertex set $Z$ with $|Z|\le(12/5)\lambda(D)$.
\end{enumerate}
\end{theorem}

The first statement specializes the cited packing theorem to directed cycles, for which the required uncrossability and oracle properties are supplied by the cited propositions. The second is the bound against the fractional optimum in the primal-dual analysis; this formulation and its combination with the first bound are also recorded in~\cite[Section~6.2 and Corollary~41]{STV}. The statements concern the actual sets computed, which is needed for the constructive algorithm stated in Lemma \ref{lem:planar-packing} and proved below.

\begin{proof}[Proof of Lemma~\ref{lem:planar-packing}]
Compute $\mathcal P$ and $Z$ by Theorem~\ref{ext:planar-tools}(i)--(ii). Their guarantees give $|Z|\le(12/5)\lambda(D)\le12|\mathcal P|$. Both computations take polynomial time.
\end{proof}

We are now ready to prove the main result of the section.
\begin{proof}[Proof of Theorem \ref{thm:planar-s4}]
After singleton preprocessing, form $D$ as in Theorem~\ref{thm:s4} and retain the list of $\eta$ unchosen conflicts. The underlying graph of $D$ is a subgraph of the planar host. Apply Lemma~\ref{lem:planar-packing} to obtain a packing
$\mathcal P$ of vertex-disjoint directed cycles and a
directed feedback vertex set $Z$ satisfying
$|Z|\le12|\mathcal P|$.

If $|\mathcal P|\ge\eta+2$, choose one of the two equalities arbitrarily for each unchosen conflict. Choose $\eta+2$ cycles from $\mathcal P$. At each vertex
on these cycles, select its incoming cycle arc; at every
remaining vertex, select an arbitrary incoming arc.
The resulting selector has at least $\eta+2$ weak
components, since each weak component contains exactly
one directed cycle. Adding the at most $\eta$ chosen equality edges leaves at least two components. This yields a solution for the $\EqSel(D,Q)$ instance for any choice of $Q$. No enumeration of the equality choices is needed in this branch.

Otherwise $|Z|\le12(\eta+1)$. The path decomposition from Lemma~\ref{lem:imbalance} has width
\[
 w\le\iota(D)+36(\eta+1).
\]
Let $T$ contain all host endpoints of the unchosen conflicts, so $|T|\le4\eta$. Expand each bag by predecessors as in Lemma~\ref{lem:pathdp} and add $T$ to every bag. This covers every chosen ternary constraint and every original unchosen constraint $(f(a)=f(b))\lor(f(c)=f(d))$, with no need to branch on equalities. The interval property is preserved. The augmented bag size is at most
\[
 3(w+1)+4\eta
 \le3\iota(D)+112\eta+111
 \le31s_4(G)+111,
\]
where the last inequality uses \(\iota(D)\le s_4(G)\) and \(\eta\le s_4(G)/4\).
For each unchosen conflict $\{ab,cd\}$, assign its constraint
to a bag containing all its endpoints. At that bag, discard
every state whose coloring $\sigma$ violates
$(\sigma(a)=\sigma(b))\lor(\sigma(c)=\sigma(d))$.
The states and all other table operations remain unchanged.
The Boolean dynamic program of Lemma~\ref{lem:pathdp}, with its used-color flag, therefore runs in $2^{O(s_4(G))}n^{O(1)}$ time and recovers a cut. All other operations are polynomial. Nonlocal conflict endpoints are handled in the bags; the algorithm does not require adding conflict or equality edges to the planar graph passed to the packing routine.
\end{proof}

The same direct treatment of extra constraints removes repeated dynamic-programming calls in the general-host algorithm: use Lemma~\ref{lem:constructive-packing} with target $\eta+2$ and add the at most $4\eta$ endpoints to the decomposition. The equality formulation remains useful for the general XP theorem and for stating the selector result.

\subsection{A tower-bounded dependence for decision}
\label{app:fpt-decision}
For decision, we can avoid the packing approximation
by using the tower-bounded choice of $F$ in
Theorem~\ref{ext:cycle-tools}(i).
Run the algorithm in part~(iii) with budget $F(h+2)$.
If no such feedback set exists, part~(i) guarantees
$h+2$ vertex-disjoint directed cycles, so
Lemma~\ref{lem:many} permits acceptance.
Otherwise, use the returned set in
Lemmas~\ref{lem:imbalance} and~\ref{lem:pathdp}.
The resulting algorithm has a tower-bounded
dependence on $h+\iota(D)$, without using $K(t)$.
\section{Proofs of the degree-five hardness results}
\label{app:degree-five}

This appendix proves both parts of
Theorem~\ref{thm:degree-five}.
We first introduce the satisfiability framework required for our hardness proofs. 
We then describe the rigid octahedron used to
represent variable occurrences and give the two
reductions. Throughout, we use the equivalent
formulation of a CF-cut as a nonconstant feasible
binary coloring.

\subsection{Satisfiability framework}
For a Boolean relation $S\subseteq\{0,1\}^r$,
an \emph{$S$-constraint} is an expression
$S(z_1,\ldots,z_r)$ whose arguments are Boolean
variables or, when permitted, Boolean constants.
Arguments may repeat. The constraint is satisfied
when the tuple of argument values belongs to $S$.

We use the following two source problems:
\begin{definition}[$\operatorname{SAT}^*(S)$ and $\operatorname{SAT}_c(S)$]
For a fixed Boolean relation $S \subseteq\{0,1\}^r$, the $\operatorname{SAT}^*(S)$ problem on a set of variables $(z_1, z_2 \ldots z_n)$, asks whether a conjunction of $S$-constraints, without constants, has a satisfying assignment different from the all 0 and all 1 assignments. 

The $\operatorname{SAT}_c(S)$ problem, on a set of variables $(z_1, z_2 \ldots z_n)$, asks whether a conjunction of $S$-constraints, has a satisfying assignment when constants zero and one are allowed in the constraints. 
\end{definition}

The following hardness result from \cite{CH}, stated for a single relation, is what will be used.
\begin{theorem}[{Creignou--H\'ebrard~\cite[Propositions~4.7--4.8]{CH}}]\label{ext:satisfiability}
If a Boolean relation $S$ is neither Horn, dual-Horn, affine, nor bijunctive, then both $\operatorname{SAT}^*(S)$ and $\operatorname{SAT}_c(S)$ are NP-complete.
\end{theorem}

The assertion for $\operatorname{SAT}_c(S)$ is Schaefer's theorem \cite{Schaefer1978}, as stated in Proposition~4.8 of the cited paper; the NP-completeness of the nonconstant variant $\operatorname{SAT}^*(S)$ is in Proposition~4.7. We recall that Horn relations are closed under coordinatewise conjunction, dual-Horn relations under coordinatewise disjunction, affine relations under the coordinatewise sum modulo two of three tuples, and bijunctive relations under coordinatewise majority of three tuples. These necessary closure properties suffice to check our use of the criterion. For full definitions of the above, the reader is referred to \cite{CH}. 

For the reductions, recall $R$ from Equation~\eqref{eq:R}. 
Its failures of the four closure properties are
\begin{align*}
 011\wedge110&=010,&001\vee100&=101,\\
 000\oplus001\oplus011&=010,&
 \operatorname{Maj}(000,011,110)&=010.
\end{align*}
All operands belong to $R$ and the displayed results do not. Thus by Theorem~\ref{ext:satisfiability} we have the following.

\begin{theorem}\label{thm:SATR}
    For the relation $R$ as defined above, both $\operatorname{SAT}^*(R)$ and $\operatorname{SAT}_c(R)$ are NP-complete.
\end{theorem}

In the context of CF-cuts, the requirement to use both colors in $\operatorname{SAT}^*(R)$ is necessary: either constant assignment satisfies every constraint when constants are absent. Section~\ref{sec:locality} uses the version $\operatorname{SAT}_c(R)$ with constants.

The \emph{primal graph} of a formula is the graph with variables as vertices, and two variables are adjacent if and only if they appear together in a clause.

\begin{observation}\label{obs:satr-connected}
    When reducing from \(\operatorname{SAT}^*(R)\), we may assume that the primal graph of the input formula is connected.
\end{observation} 
\begin{proof}
    If not, we connect its components using tautologies $R(x,x,y)$, which do not change any assignment. An unused declared variable can likewise be included by such a tautology; it must not simply be dropped when it could witness nonconstancy. Instances with fewer than two declared variables are trivial and can be handled separately. Thus all variables have occurrences, and the graph joining variables that share a constraint is connected.
\end{proof}

\subsection{The rigid octahedron}\label{app:localfive}

A graph with specified conflicts is \emph{rigid} if
every feasible coloring is constant; a copy is called
a \emph{rigid block}. We use the following rigid
octahedron with exactly $1$-regular conflicts to
represent the value of a source variable.
\begin{equation}\label{eq:rigidU}
 U=K_6-\{03,15,24\},\qquad
 (01,02),(12,14),(23,25),(13,34),(04,45),(05,35).
\end{equation}
The pairs listed after the graph represent the edge conflict pairs. It can be noted that $\widehat G$ is exactly 1-regular and the conflicts are all local. This was shown to be rigid in \cite{KK} and we also provide a short proof %
below.
For exactly $2$-regular conflicts, there is a smaller rigid block: take a host triangle and let its three edges form a triangle in the conflict graph. Every nonconstant coloring of a triangle cuts two edges, and those two edges conflict. Hence the block is rigid.

\begin{lemma}\label{lem:U}
The octahedron with conflicts specified in~\eqref{eq:rigidU} is rigid. Its conflicts are local and exactly $1$-regular.
\end{lemma}
\begin{proof}
The twelve edges of $U=K_6-\{03,15,24\}$ occur exactly once among the six listed pairs. Pair $i$ is centered at vertex $i$ for $i=0,1,2$, the fourth pair is centered at $3$, the fifth at $4$, and the last at $5$.

Write $x_i$ for the color of vertex $i$ and, by complement symmetry, fix $x_0=0$. The pair centered at zero gives $(x_1,x_2)\in\{00,01,10\}$. If $(x_1,x_2)=00$ and $x_3=1$, the pair at $3$ forces $x_4=1$, and the pair at $4$ forces $x_5=1$. This contradicts the pair at $2$. Hence $x_3=0$, after which the pairs at $5$ and $4$ force $x_5=x_4=0$. This is the all-zero coloring.

If $(x_1,x_2)=01$, the pair at $1$ forces $x_4=0$, the pair at $3$ forces $x_3=0$, and the pair at $5$ forces $x_5=0$. The pair at $2$ then fails. Finally, if $(x_1,x_2)=10$, the pairs at $1,4,5$ force successively $x_4=x_5=x_3=1$, again contradicting the pair at $2$. Only the all-zero coloring remains, and complementation gives the other constant coloring.
\end{proof}

\subsection{Proof of Theorem~\ref{thm:degree-five}(i)}
\label{app:localfiveproof}

\begin{proof}%
Observation~\ref{obs:satr-connected} allows us to reduce from the connected version of
$\operatorname{SAT}^*(R)$, in which every
variable occurs in some constraint. We construct a CF-cut
instance with host graph $G$.
We use the equivalent coloring formulation from
Section~\ref{sec:preliminaries}: the constructed instance
has a CF-cut if and only if it admits a nonconstant
feasible $\bits$-coloring.
Given a $\operatorname{SAT}^*(R)$ instance, 
for each variable $x$ with $t_x$  occurrences, create a chain $U_x^1,\ldots,U_x^{t_x}$ of disjoint copies of the rigid octahedron. Associate one copy with each occurrence of the variable $x$. Repeated occurrences within one clause receive different copies. We call the six vertices of each copy its \emph{ports}.
Each port receives at most one attachment edge; a port
with no attachment edge is \emph{unused}.

Between consecutive copies introduce a fresh vertex $w$. Join it to two distinct vertices of the first copy, pairing those two edges with conflicts, and to two distinct vertices of the second copy, pairing those two edges. These conflict pairs force $w$ to agree with each monochromatic copy. Hence every copy in the chain and every intervening connector have a common variable color. 

\paragraph*{Clause vertices.}
For a source constraint $R(a,b,c)$, add a fresh vertex $z$. Join it to one unused port in its $a$-occurrence copy, one unused port in its $c$-occurrence copy, and two distinct unused ports in its $b$-occurrence copy. Pair the two $b$-edges; this forces $z$ to have the
color of its $b$-occurrence block. Pair the $a$-edge
and $c$-edge; the resulting constraint is precisely
$R$ applied to the three occurrence-block colors.

\begin{figure}[tb]
\centering
\begin{tikzpicture}[x=1cm,y=1cm,every node/.style={text=cfcHost,font=\footnotesize}]
 \path[use as bounding box] (0,0) rectangle (13.8,6.35);
 \node[cfc panel] at (.05,6.13) {(a) Occurrence chain};
 \CFCompactBlock{X1}{1}{5.1}{1}{$U_x^1$}
 \CFCompactBlock{X2}{6.9}{5.1}{1}{$U_x^2$}
 \CFCompactBlock{X3}{12.8}{5.1}{1}{$U_x^3$}
 \CFCompactEquality{w1}{3.95,5.1}{X1}{X2}
 \CFCompactEquality{w2}{9.85,5.1}{X2}{X3}
 \node[below=3pt] at (w1) {$w_1$};
 \node[below=3pt] at (w2) {$w_2$};
 \node[cfc panel] at (.05,3.73) {(b) Clause connector for $R(a,b,c)$};
 \CFCompactBlock{A}{1}{2.65}{1}{$U_a$}
 \CFCompactBlock{B}{6.9}{2.95}{1}{$U_b$}
 \CFCompactBlock{C}{12.8}{2.65}{-1}{$U_c$}
 \coordinate (z) at (6.9,.8);
 \draw[cfc edge] (A-4)--node[pos=.47,below left=1pt] {$e_a$}(z);
 \draw[cfc edge] (C-4)--node[pos=.47,below right=1pt] {$e_c$}(z);
 \draw[cfc edge] (B-4)--node[pos=.40,right=2pt] {$g_4$}(z);
 \draw[cfc edge] (B-5)--node[pos=.40,left=2pt] {$g_5$}(z);
 \draw[cfc conflict] ($(z)!.3!(B-4)$) to[bend right=28] ($(z)!.3!(B-5)$);
 \draw[cfc conflict] ($(z)!.12!(A-4)$) .. controls ($(z)+(-.85,-.65)$) and ($(z)+(.85,-.65)$) .. ($(z)!.12!(C-4)$);
 \node[cfc vertex] at (z) {};
 \node at ($(z)+(.35,-.12)$) {$z$};
\node[anchor=west] at (7.8,.32) {$f(z)=f(U_b)$};
\end{tikzpicture}
\caption{Gadgets for Theorem~\ref{thm:degree-five}(i).
Hexagons are rigid octahedra; internal edges and
conflicts are omitted. Dashed red links join
conflicting edges. Equality connectors synchronize
occurrences. Writing $f(U_x)$ for the common color
of block $U_x$, the pair $\{g_4,g_5\}$ forces
$f(z)=f(U_b)$, after which $\{e_a,e_c\}$ enforces
$R(f(U_a),f(U_b),f(U_c))$.}
\label{fig:compact-local-five}
\end{figure}
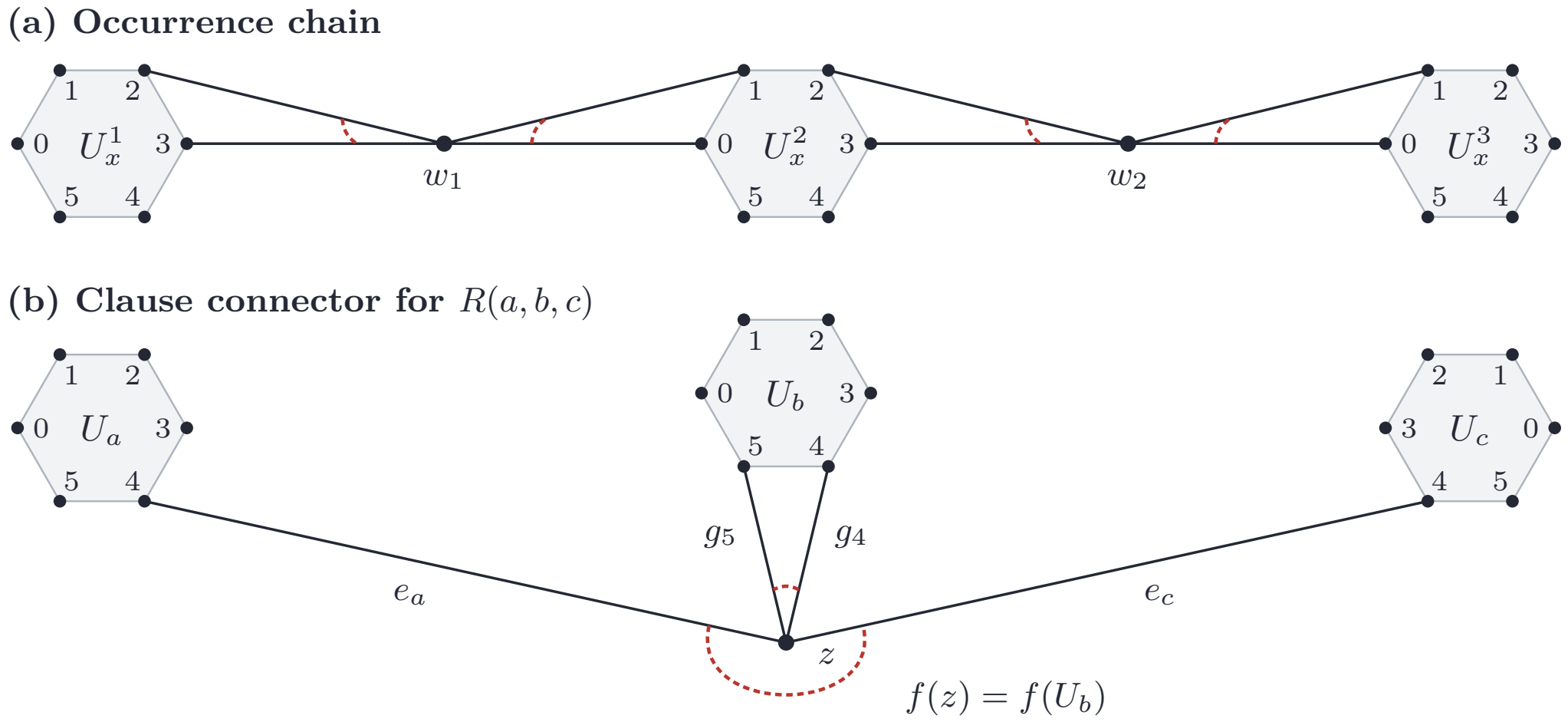

All attachment vertices can be chosen without reuse. A copy needs two ports for each neighboring chain connector, at most four in total, and at most two for its own clause occurrence. It has six vertices, each initially of degree four. Assign these incidences injectively to its vertices. Every block vertex consequently has degree at most five, the clause connectors ensure every block has at least one vertex with degree five, and every connector has degree four. Each new edge is introduced in exactly one of the specified pairs. Internal block edges retain their original mates. All pairs are local and all edges of $G$ participate in exactly one conflict. Fresh connectors and distinct ports ensure a simple graph. The connectivity of the primal graph of the source ensures the constructed graph $G$ is connected.

\paragraph*{Correctness of the reduction.}
A nonconstant satisfying assignment of the source $\operatorname{SAT}^*(R)$ 
instance extends by giving each occurrence block its
variable value, every equality connector that same
value, and every clause vertex its middle-variable
value. This gives a nonconstant feasible coloring.

Conversely, every feasible coloring is constant on
each variable chain, and the clause pairs ensure that
the resulting variable values satisfy every source
constraint. Every equality connector has its chain's
color, and every clause vertex has its middle-variable
block's color. Hence a constant source assignment
would make the entire graph monochromatic. A
nonconstant feasible coloring therefore yields a
nonconstant satisfying source assignment.
This proves NP-hardness under all stated restrictions, and the problem belongs to NP.
\end{proof}

\subsection{Proof of Theorem~\ref{thm:degree-five}(ii)}
\label{app:exactfive}

We start with the construction used for Theorem~\ref{thm:degree-five}(i), but arrange the occurrence blocks cyclically. Every vertex whose degree is still four can then be joined to a vertex that is already forced to have the same color. These new edges never cross a feasible cut, and hence can be paired without changing the solutions.

\begin{proof}
By Observation~\ref{obs:satr-connected}, we may start
with a connected instance of $\operatorname{SAT}^*(R)$
in which every variable occurs. Let $\ell$ be its
number of constraints. Duplicate each constraint,
obtaining $2\ell$ constraints and $6\ell$ coordinate
occurrences. This preserves all satisfying assignments
and ensures that every variable has at least two
occurrences.

\paragraph*{Occurrence blocks and equality vertices.}
Create a separate copy of the rigid octahedron $U$ for every occurrence of a variable, including different occurrences of the same variable in a clause. Label the vertices in an occurrence block $U_i$ by $0_i,\ldots,5_i$, according to Equation~\eqref{eq:rigidU}. For each variable, arrange its blocks in a cyclic order. Between $U_i$ and its successor $U_{i+1}$, introduce a fresh vertex $w_i$. Here indices are interpreted cyclically within the
occurrence blocks of each variable. Add the four edges from $w_i$ to $2_i,3_i,0_{i+1},1_{i+1}$, with conflict pairs
\[
 \{w_i2_i,w_i3_i\},\qquad
 \{w_i0_{i+1},w_i1_{i+1}\}.
\]
The first pair forces $w_i$ to have the color of $U_i$, and the second forces it to have the color of $U_{i+1}$. All occurrences of a variable are therefore synchronized. Each block uses ports 0 and 1 for its incoming connector and ports 2 and 3 for its outgoing connector. When a variable has two occurrences, the two connectors are distinct vertices, so the construction is still simple.

For every copied constraint $R(a,b,c)$, create a clause vertex $z$. Join $z$ to ports 4 and 5 in the middle-occurrence block and pair those edges. Join $z$ to port 4 in each outer-occurrence block and pair these two edges. As in Theorem~\ref{thm:degree-five}(i), the first pair
forces $z$ to have the color of the middle-occurrence
block, and the second enforces $R$ on the three
occurrence-block colors.

\begin{claim}\label{cl:five-deficit}
At this stage, the vertices of degree four are exactly the equality vertices $w_i$, the clause vertices $z$, and port 5 in each outer-occurrence block. Every other vertex has degree five.
\end{claim}
\begin{claimproof}
Each octahedron vertex has four internal neighbors. Ports 0 through 3 each receive one synchronization edge. In a middle-occurrence block, ports 4 and 5 each receive one clause edge. In an outer-occurrence block, only port 4 receives a clause edge. Every equality vertex and every clause vertex has four incident edges by construction.
\end{claimproof}

\paragraph*{Completing the degrees.}
For each occurrence block $U_i$, add one edge incident with its outgoing equality vertex $w_i$. If the occurrence is an outer coordinate, add $w_i5_i$. If it is a middle coordinate, add $w_i z$, where $z$ is its clause vertex. See Figure~\ref{fig:compact-exact-five} for an illustration.

\begin{figure}[tb]
\centering
\begin{tikzpicture}[x=1cm,y=1cm,every node/.style={text=cfcHost,font=\footnotesize}]
 \path[use as bounding box] (0,0) rectangle (13.8,8.35);
 \node[cfc panel] at (.05,8.12) {(a) Cyclic synchronization};
 \CFCompactBlock{Y1}{1.05}{7.1}{1}{$U_1$}
 \CFCompactBlock{Y2}{4.95}{7.1}{1}{$U_2$}
 \CFCompactBlock{Y3}{8.85}{7.1}{1}{$U_3$}
 \CFCompactBlock{Y4}{12.75}{7.1}{1}{$U_4$}
 \CFCompactEquality{v1}{3,7.1}{Y1}{Y2}
 \CFCompactEquality{v2}{6.9,7.1}{Y2}{Y3}
 \CFCompactEquality{v3}{10.8,7.1}{Y3}{Y4}
 \node[below=3pt] at (v1) {$w_1$};
 \node[below=3pt] at (v2) {$w_2$};
 \node[below=3pt] at (v3) {$w_3$};
 \coordinate (v4) at (6.9,5.35);
 \draw[cfc edge] (v4)--(.06,5.35)--(.06,7.1)--(Y1-0);
 \draw[cfc edge] (v4)--(6.1,5.8)--(.2,5.8)--(.2,7.9)--(.5,7.9)--(Y1-1);
 \draw[cfc edge] (v4)--(13.68,5.35)--(13.68,7.9)--(13.25,7.9)--(Y4-2);
 \draw[cfc edge] (v4)--(7.7,5.8)--(13.54,5.8)--(13.54,7.1)--(Y4-3);
 \draw[cfc conflict] (6.43,5.35) to[bend right=25] (6.50,5.575);
 \draw[cfc conflict] (7.37,5.35) to[bend left=25] (7.30,5.575);
 \node[cfc vertex,label={[label distance=2pt]below:$w_4$}] at (v4) {};

 \node[cfc panel] at (.05,4.67) {(b) Degree completion};
 \node at (3.05,4.15) {Outer occurrence};
 \node at (10.35,4.15) {Middle occurrence};
 \CFCompactBlock{O}{2.9}{3.3}{1}{$U_i$}
 \CFCompactBlock{M}{10.3}{3.3}{1}{$U_i$}
 \coordinate (ol) at (1.15,3.3);
 \coordinate (or) at (4.65,3.3);
 \coordinate (ml) at (8.55,3.3);
 \coordinate (mr) at (12.05,3.3);
 \foreach \p/\l/\r in {O/ol/or,M/ml/mr}{
  \draw[cfc edge] (\p-0)--(\l)--(\p-1);
  \draw[cfc edge] (\p-2)--(\r)--(\p-3);
  \draw[cfc conflict] ($(\l)!.34!(\p-0)$) to[bend left=25] ($(\l)!.34!(\p-1)$);
  \draw[cfc conflict] ($(\r)!.34!(\p-2)$) to[bend right=25] ($(\r)!.34!(\p-3)$);
  \node[cfc vertex,label={[label distance=2pt]below:$w_{i-1}$}] at (\l) {};
  \node[cfc vertex,label={[label distance=2pt]above right:$w_i$}] at (\r) {};
 }
 \coordinate (oz) at (3.265,1.65);
 \coordinate (mz) at (10.3,1.65);
 \draw[cfc edge] (O-4)--(oz);
 \draw[cfc edge] (M-4)--(mz)--(M-5);
 \draw[cfc conflict] ($(mz)!.31!(M-4)$) to[bend right=28] ($(mz)!.31!(M-5)$);
 \draw[cfc added] (or) .. controls (5,.45) and (1.5,.45) .. (O-5);
 \draw[cfc added] (mr) .. controls (12.7,1) and (11.55,.9) .. (mz);
 \node[cfc vertex,label={[label distance=2pt]right:$z$}] at (oz) {};
 \node[cfc vertex,label={[label distance=2pt]left:$z$}] at (mz) {};
 \node[text=cfcAdded] at (3.05,.78) {add $w_i5_i$};
 \node[text=cfcAdded] at (10.35,.78) {add $w_i z$};
 \draw[cfc edge] (.3,.16)--(.85,.16);
 \node[anchor=west] at (1,.16) {host edge};
 \draw[cfc conflict] (4.55,.1) to[bend left=25] (5.1,.1);
 \node[anchor=west] at (5.25,.16) {conflict};
 \draw[cfc added] (8.45,.16)--(9,.16);
 \node[anchor=west] at (9.15,.16) {added edge};
\end{tikzpicture}
\caption{Gadget for Theorem~\ref{thm:degree-five}(ii). Occurrences are synchronized cyclically; the clause gadget is unchanged from Figure~\ref{fig:compact-local-five}. Blue edges join vertices of degree four that are already forced to have the same color. For an outer occurrence, add $w_i5_i$; for a middle occurrence, add $w_i z$. Other edges at connectors are omitted in (b). Pairing the added edges arbitrarily gives nonlocal conflicts and preserves all feasible colorings.}
\label{fig:compact-exact-five}
\end{figure}
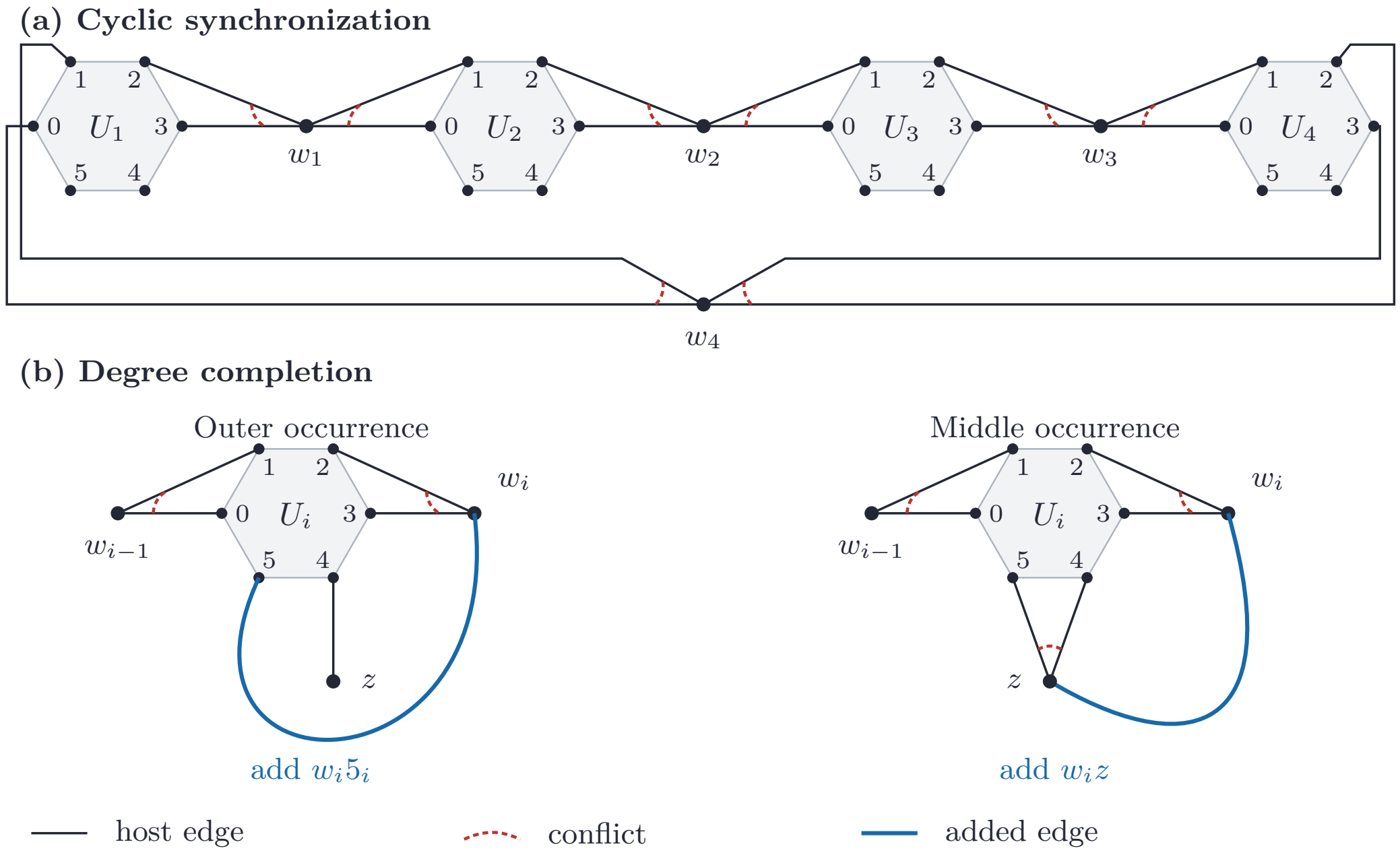

\begin{claim}\label{cl:five-completion}
The resulting graph is simple and $5$-regular. Every added edge is monochromatic in every coloring satisfying the conflicts introduced so far.
\end{claim}
\begin{claimproof}
Each equality vertex $w_i$ receives one new edge. Each unused outer port $5_i$ receives one, and each clause vertex receives one from the equality vertex
corresponding to its unique middle occurrence. By Claim~\ref{cl:five-deficit}, these are exactly the vertices that required one more edge. None of the added edges was present: $w_i$ was adjacent only to ports 2 and 3 in its own block and ports 0 and 1 in the next block; there were no equality-to-clause edges. Thus the added edges form a matching covering exactly
the vertices that previously had degree four.

The endpoints of $w_i5_i$ both have the color of $U_i$. For a middle occurrence, $w_i$ and $z$ both have the color of its variable. Thus every added edge is monochromatic.
\end{claimproof}

There are $6\ell$ occurrence blocks and hence $6\ell$ added edges. Pair these edges arbitrarily and declare each pair a conflict. Every old edge keeps its unique mate, and every new edge receives one mate. Thus the conflict graph is exactly $1$-regular. By Claim~\ref{cl:five-completion}, all the new conflicts are satisfied whenever the old ones are satisfied.

\paragraph*{Correctness.}
A nonconstant satisfying assignment colors each block and equality vertex by its variable value and each clause vertex by its middle value. The gadget arguments show that all old conflicts hold, and the added edges are monochromatic. The resulting coloring is nonconstant.

Conversely, a feasible coloring of the final instance satisfies all old conflicts. Rigidity and synchronization give a value for each source variable, and the clause pairs enforce every source constraint. 
Every equality vertex has its variable's color, and
every clause vertex has its middle variable's color.
Consequently, if all source variables had the same
value, the entire graph would be monochromatic.
A nonconstant feasible coloring therefore yields a
nonconstant satisfying source assignment.

Connectedness follows from the connected primal graph of the source. The final graph has $44\ell$ vertices, $110\ell$ edges and $55\ell$ conflict pairs, so the reduction has linear size. A proposed cut can be checked in polynomial time, proving membership in NP.
\end{proof}

The added edges in this proof form a matching in $G$. Pairing them therefore creates nonlocal conflicts. Theorem~\ref{thm:degree-five} gives two different refinements of maximum-degree-five hardness: the first imposes locality, and the second imposes exact regularity of $G$.

\section{Proofs for conflict components}
\label{app:components}

We give the propagation proof used in
Section~\ref{sec:degree-hardness}, followed by
the deferred hardness and classification proofs.
We conclude with the detailed proof of the
exceptional-components algorithm.

\subsection{Proof of Lemma~\ref{lem:majority-propagation}}
\label{app:majority-propagation}

\begin{proof}[Proof of Lemma~\ref{lem:majority-propagation}]
No vertex changes more than once. If an output changes to one, its two one-colored inputs remain one thereafter. When no further change is possible, every output that is still zero has at most one one-colored input. Thus all equations at the remaining vertices are satisfied. 

For the implementation, let $S$ be the set of vertices
whose colors are not fixed. For each $v\in S$, maintain
the number of its three specified inputs currently
colored one. For each vertex $x$, also store the list of equations
with output in $S$ in which $x$ occurs as an input.

Initialize the counters and enqueue every vertex of
$S$ with at least two one-colored inputs. When a vertex
$v$ is dequeued, change its color to one and update the
counters of the equations having $v$ as an input.
Enqueue an output when its counter first reaches two,
marking vertices when enqueued so that each is
enqueued at most once.

There are $3|S|$ input occurrences in total. Each
vertex is processed at most once, and each input
occurrence is examined at most once during these
updates. Including initialization, the running time
is $O(n)$. Equations at fixed vertices, if any, are not asserted by the lemma and must be checked separately.
\end{proof}

\subsection{Proof of Theorem~\ref{thm:two-four}}
\label{app:two-four}

\begin{proof}[Proof of Theorem~\ref{thm:two-four}]
By Observation~\ref{obs:satr-connected}, we reduce
from $\operatorname{SAT}^*(R)$ restricted to instances
with connected primal graph in which
every variable occurs in a constraint. 
We construct a CF-cut instance $(G,\widehat G)$ where $G$ is 4-regular, and all the components of $\widehat G$ are $C_3$'s and $C_4$'s.
We use the equivalent coloring formulation from
Section~\ref{sec:preliminaries}: the constructed instance
has a CF-cut if and only if it admits a nonconstant
feasible coloring $f: V(G) \rightarrow \bits$.
First we describe the graph and conflicts, and then prove that the construction has the required properties and preserves the solutions. 
Replace every clause in the \(\operatorname{SAT}^*(R)\) instance by four identical copies; this preserves all satisfying assignments. Each variable occurrence gets a separate rigid triangle, even when symbols repeat in a clause. Its three vertices are denoted by $t_0,t_1,t_2$. For a rigid triangle $A$, write $f(A)$ for the
common color of its vertices. Internal edges of each triangle form their own conflict component $C_3$.

\paragraph*{Equality connectors.}
Let $A,B$ be two rigid triangles. Add a fresh vertex $z$, two edges $a_1,a_2$ from $z$ to distinct vertices in the first triangle, and two edges $b_1,b_2$ to distinct vertices in the second. On these four edges of $G$ put the conflict cycle
\[
 a_1a_2b_1b_2a_1.
\]
The conflict $a_1a_2$ forces $f(z)=f(A)$: if $f(z)\ne f(A)$, both edges cross. Similarly $b_1b_2$ forces $f(z)=f(B)$, yielding $f(z)=f(A) = f(B)$. Conversely, if $f(z)=f(A) = f(B)$, none of the four edges crosses and all four conflicts hold.

For each variable, order its occurrence triangles cyclically, and put one equality connector between each consecutive pair. There are at least four occurrences, because constraints were quadrupled. The connector from a triangle to its successor uses $t_0,t_2$ in that triangle and $t_0,t_1$ in the successor. Each triangle consequently receives one external incidence at each of $t_1,t_2$ and two at $t_0$. Including internal triangle edges, the degrees at this point are $(4,3,3)$. Every triangle of a variable and every intervening connector have one common color.

\paragraph*{Clause connectors.}
For a copied clause $R(a,b,c)$, use its three distinct occurrence triangles $A,B,C$ and add a fresh vertex $z$. Join $z$ to $t_1$ of $A$ by an edge $e_A$, to $t_1$ of $C$ by $e_C$, and to $t_1,t_2$ of $B$ by $e_{B,1},e_{B,2}$. Put the conflict cycle
\begin{equation}\label{eq:clausecycle}
 e_Ae_Ce_{B,2}e_{B,1}e_A.
\end{equation}
The pair $e_{B,1} e_{B,2}$ forces $f(z)=f(B)$. Once this holds, both edges to $B$ are monochromatic, so their other conflicts are automatic. The remaining pair $e_Ae_C$ is respected exactly when $(f(A)=f(B))\lor(f(B)=f(C))$. This enforces the source relation
$R(f(A),f(B),f(C))$. The clause vertex has degree four. Both free ports of the middle triangle are now filled, and the outer triangles each have the single free port $t_2$.

\begin{figure}[tb]
\centering
\begin{tikzpicture}
\node[block] (A) at (0,1.0) {$A$};
\node[block] (B) at (0,-1.0) {$B$};
\node[vertex] (z) at (2,0) {$z$};
\draw (z)--($(A.east)+(0,.12)$) node[midway,above] {$a_1$};
\draw (z)--($(A.east)+(0,-.12)$) node[midway,below] {$a_2$};
\draw (z)--($(B.east)+(0,.12)$) node[midway,above] {$b_1$};
\draw (z)--($(B.east)+(0,-.12)$) node[midway,below] {$b_2$};
\foreach \n/\xx/\yy in {a_1/3.1/1,a_2/4.5/1,b_1/4.5/-1,b_2/3.1/-1}
 \node[vertex] (\n) at (\xx,\yy) {$\n$};
\draw[conflict] (a_1)--(a_2)--(b_1)--(b_2)--(a_1);
\node at (2.2,-1.7) {(a) equality};
\begin{scope}[xshift=6.3cm]
\node[block] (A) at (0,1) {$A$};
\node[block] (C) at (2.1,1) {$C$};
\node[block] (B) at (1.05,-1) {$B$};
\node[vertex] (z) at (1.05,0) {$z$};

\draw (z)--(A) node[midway,left] {$e_A$};
\draw (z)--(C) node[midway,right] {$e_C$};
\draw (z)--($(B.north)+(-.15,0)$)
  node[midway,left] {$e_{B,1}$};
\draw (z)--($(B.north)+(.15,0)$)
  node[midway,right] {$e_{B,2}$};

\node[vertex] (eA) at (3.2,1) {$e_A$};
\node[vertex] (eC) at (4.7,1) {$e_C$};
\node[vertex] (eBtwo) at (4.7,-1) {$e_{B,2}$};
\node[vertex] (eBone) at (3.2,-1) {$e_{B,1}$};

\draw[conflict]
  (eA)--(eC)--(eBtwo)--(eBone)--(eA);

\node at (2.4,-1.7) {(b) clause};
\end{scope}
\end{tikzpicture}
\caption{The two degree-four connectors. A box represents a rigid monochromatic block; two attachments to a box use distinct vertices. The edges of $G$ appear on the left of each panel, and the prescribed conflict four-cycle on the right. Equality forces $f(A)=f(B)=f(z)$; the clause connector
forces $f(z)=f(B)$ and then enforces
$R(f(A),f(B),f(C))$.}
\label{fig:degree-two}
\end{figure}
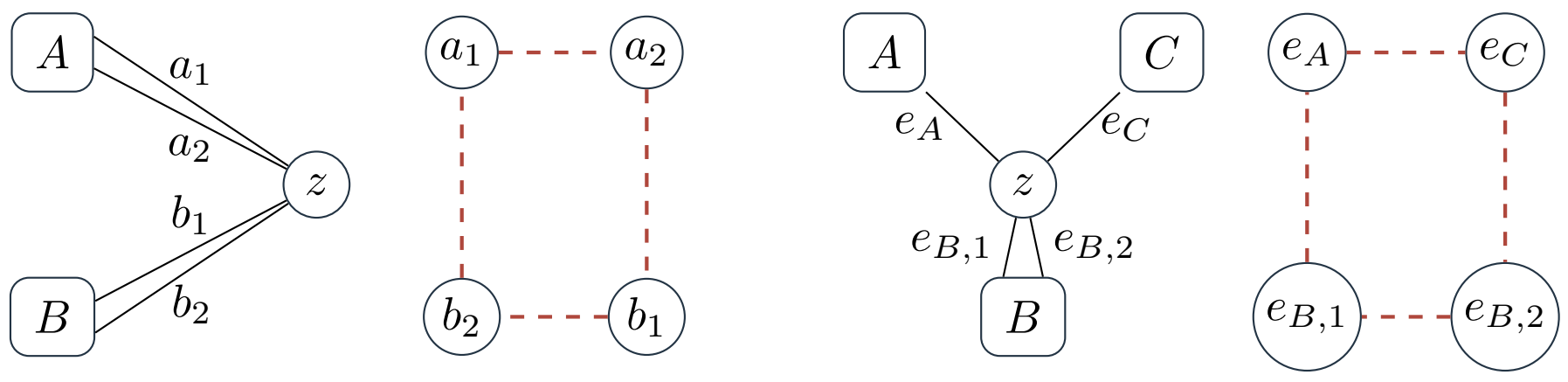

\paragraph*{Completing degree four with padding stars.}
For a fixed variable, the number of outer-coordinate occurrences is divisible by four. Group its unused $t_2$ ports in groups of four. For each group add a fresh vertex $w$ adjacent to its four ports, and let those four edges of $G$ form a conflict cycle $C_4$. The four neighbors of $w$ already have the same variable color. If $w$ differed from it, all four edges would cross and violate the conflict cycle. If it agrees, none crosses. Thus this padding forces the intended color and introduces no new assignment. All ports are now filled and every vertex has degree exactly four.

\paragraph*{Degrees, conflicts and connectedness.}
Every internal triangle edge belongs only to that triangle's conflict component. Every other edge belongs to exactly one connector or padding star and hence to exactly one of the four-cycles specified above. Thus the conflict graph is precisely a disjoint union of $C_3$ and $C_4$, so every edge of $G$ has exactly two conflicts. Conflicts within a triangle are local, and 
all four edges associated with each equality connector,
clause connector, or padding star share its center, so the entire conflict system is local. Distinct occurrence triangles and fresh connector vertices ensure simplicity, including repeated variables within a constraint. Each variable's triangles are connected by its synchronization cycle; clause vertices connect their variable blocks. Connectedness of the source therefore implies connectedness of the host.

If the source instance of \(\operatorname{SAT}^*(R)\) has $\ell$ constraints before quadrupling, the construction has $12\ell$ triangles, $12\ell$ equality connectors, $4\ell$ clause connectors, and $2\ell$ padding vertices. Hence it has $54\ell$ vertices and $108\ell$ edges. The conflict graph has $12\ell$ triangle components and $18\ell$ four-cycle components. These counts also confirm polynomial size and both regularity conditions.

\paragraph*{Correctness of the reduction.}
A satisfying  assignment of the source \(\operatorname{SAT}^*(R)\) instance colors each variable's triangles and equality connectors with its value, each clause vertex with its middle value, and each padding vertex with its variable value. The gadget arguments show that all conflicts hold. If the source is nonconstant, triangles of differently colored variables certify a nonconstant target coloring.

Conversely, any feasible target coloring makes each triangle monochromatic. Equality connectors synchronize all occurrences of each variable. Clause connectors then enforce every source relation. All equality and padding vertices are forced to variable colors, and every clause vertex is forced to its middle variable's color. Therefore a constant assignment on all source variables would make the whole target constant. A nonconstant feasible target coloring yields a nonconstant satisfying source assignment. This proves equivalence and NP-hardness. A nonconstant vertex coloring is a polynomially checkable certificate, giving membership in NP.
\end{proof}

\subsection{Higher regular degrees}
\label{app:higherdegree}

We extend the construction of
Theorem~\ref{thm:two-four} to higher regular degrees
by adding edges between vertices that are forced
to have the same color.

\begin{corollary}\label{cor:higher}
For each fixed integer $D\ge4$, \CF{} is NP-complete on simple connected $D$-regular hosts in each of the following regimes:
\begin{enumerate}[(i)]
\item local conflicts with $\Delta(\widehat G)\le2$;
\item exactly $2$-regular conflicts, without a locality requirement.
\end{enumerate}
For $D=4$, Theorem~\ref{thm:two-four} satisfies both restrictions simultaneously.
\end{corollary}
\begin{proof}
The case $D=4$ follows from Theorem~\ref{thm:two-four},
so assume $D\ge5$.
Replace every source clause by $2D$ identical copies,
keeping the same variables, and apply that theorem's
construction, including its fourfold clause repetition.
For each source variable $x$, group together its
occurrence-triangle vertices, equality connectors,
padding vertices, and clause vertices whose middle
variable is $x$. Call this group the class of $x$.
Every feasible coloring is constant on each class.
The additional repetition multiplies each class size
by $2D$, so its resulting size $N_x$ is even and
at least $2D$.
Clause repetition preserves all satisfying assignments
and, for fixed $D$, increases the construction size
by only a constant factor.

Add $D-4$ pairwise edge-disjoint perfect matchings inside each class, avoiding every existing edge. To justify their existence, before the $(j+1)$st added matching, where $0\le j\le D-5$, every vertex has at most $4+j\le D-1$ neighbors already present inside its class. The graph of available pairs therefore has minimum degree at least $N_x-D\ge N_x/2$. 
A graph of even order $N_x$ with minimum degree
at least $N_x/2$ has a perfect matching.%
\footnote{Let $M$ be a maximum matching.
If it is not perfect, choose two unmatched vertices
$u,v$. Every neighbor of either vertex is matched.
For each $ab\in M$, at most two of the edges
$ua,ub,va,vb$ can be present; otherwise $ab$
could be replaced by two disjoint edges matching
$u$ and $v$, enlarging $M$.
Hence
$d(u)+d(v)\le2|M|\le N_x-2$,
contradicting the minimum-degree assumption.}
The desired matching at each step can be found using a standard algorithm \cite{Edmonds1965}.

After these additions the host is simple, connected, and $D$-regular. Every added edge is monochromatic in all feasible colorings of the old instance. Leaving them unpaired in $\widehat G$ proves (i): the old local conflicts remain and new conflict vertices are isolated. Feasible colorings are unchanged in either direction.

For (ii), group the added edges arbitrarily into triples and make each triple a conflict triangle. If the amplified source before the construction has $\ell$ clauses, the original host has $54\ell$ vertices, so the number of added edges is $27\ell(D-4)$, divisible by three. Original conflict vertices keep degree two, and each new edge now has exactly two conflicts. These new conflicts are harmless because their edges are forced monochromatic. They need not be local. This proves (ii) while keeping the two promised regimes separate.
\end{proof}

\subsection{Proof of Theorem~\ref{thm:component-dichotomy}}
\label{app:component-dichotomy}

\begin{proof}[Proof of Theorem~\ref{thm:component-dichotomy}]
Under locality and the host-degree bound, only four clique sizes can occur. Indeed, three pairwise-intersecting edges of $G$ form a triangle or share an endpoint, as in Observation~\ref{obs:triangle-types}. No fourth distinct edge intersects all three edges of a triangle: an edge through one triangle vertex must meet the opposite edge, making it another triangle edge, and an edge with no triangle endpoint meets none. If the three edges form a star, a fourth edge meeting all three must contain its center, since otherwise its two endpoints would have to meet three distinct leaves. Thus four pairwise-intersecting edges share an endpoint. Any further edge meeting all four must share that endpoint as well. The bound $\Delta(G)\le4$ therefore excludes conflict cliques larger than $K_4$, and each conflict $K_4$ is supported on a four-edge host star.

The polynomial cases follow from
Theorems~\ref{thm:triangles} and~\ref{thm:density},
together with a singleton-cut counting argument.
For hardness, we retain the host from the reduction
in Theorem~\ref{thm:two-four} and describe only the
modifications to its conflicts, verifying that they
preserve feasible colorings.

\paragraph*{Polynomial cases.}
If neither $K_2$ nor $K_4$ is allowed, every component is $K_1$ or $K_3$, and Theorem~\ref{thm:triangles} applies. It remains to consider the case $K_3\notin\mathcal A$. If the input has no $K_4$ component, its conflicts form a matching. Since $m\le2n$, Theorem~\ref{thm:density} applies.

Suppose a $K_4$ component occurs. We show that some single-vertex cut is feasible. Otherwise, every vertex of $G$ centers a conflict, and hence centers a $K_2$ component or a $K_4$ star component. Choose one such component at each vertex. These chosen components are distinct: a local $K_2$ has a unique common host endpoint, and a $K_4$ star also has a unique center. Components partition edges of $G$, so the chosen components have disjoint edge sets and use at least two edges of $G$ each.

Let $z$ be the center of any $K_4$ component. Its four edges exhaust the incident edges of $z$. Consequently $z$ cannot center any other component and must choose that $K_4$. The selected components use at least $2(n-1)+4=2n+2$ edges of $G$, contradicting $m\le2n$. Thus testing the single-vertex cuts finds a solution. This completes the polynomial cases.

\paragraph*{Hardness with $K_2$ and $K_3$.}
Keep the host and every rigid-triangle component of
the reduction in Theorem~\ref{thm:two-four}. Replace
each conflict four-cycle by a perfect matching of
that cycle, chosen as follows.

At an equality connector retain $\{a_1,a_2\}$ and
$\{b_1,b_2\}$. These already force
$f(z)=f(A)=f(B)$, so every deleted conflict is
automatic. At a clause connector retain
$\{e_A,e_C\}$ and $\{e_{B,1},e_{B,2}\}$.
The latter forces $f(z)=f(B)$, making both edges
to $B$ monochromatic, and the former enforces
$R(f(A),f(B),f(C))$. Each deleted conflict contains
one of the monochromatic edges to $B$.
At a padding star retain any perfect matching of
its conflict four-cycle. All four leaves already
have the same color, and a retained pair forces
the center to that color.

Every coloring satisfying the retained constraints therefore satisfies all removed constraints; the converse is immediate. Feasible nonconstant colorings are unchanged. The host remains simple, connected, and exactly $4$-regular, and the conflict graph consists only of local $K_2$ and $K_3$ components.

\paragraph*{Hardness with $K_3$ and $K_4$.}
Again start from Theorem~\ref{thm:two-four}, but complete each conflict four-cycle to a $K_4$ on the same four edges of $G$. Keep the rigid-triangle conflict components unchanged. Every completed component is supported on a four-edge host star, so all added conflicts are local.

The added conflicts are redundant in every feasible coloring of the original construction. At an equality connector all four edges are monochromatic, since its center and both attached rigid blocks have the same color. 
At a clause connector, the edges $e_{B,1}$ and
$e_{B,2}$ are monochromatic. The only pair avoiding
these edges is $\{e_A,e_C\}$, which already conflicts.
Every new conflict therefore contains a
monochromatic edge.
At a padding star the center and all four leaves have the same variable color, so again all four edges are monochromatic. It follows that every original feasible coloring satisfies the completed components. The reverse implication holds because completion only adds constraints.

Thus completion preserves feasible nonconstant colorings exactly. The host is unchanged and remains simple, connected, and exactly $4$-regular. Its conflict graph has only $K_3$ and $K_4$ components. This proves the second hardness case. Both constructions have polynomial size, and
membership in NP follows by checking that a proposed
nonconstant vertex coloring satisfies every conflict.
\end{proof}

\subsection{Proof of Theorem~\ref{thm:exceptions}}
\label{app:majority-extension}

\begin{proof}[Proof of Theorem~\ref{thm:exceptions}]
Test single-vertex cuts first and return one if feasible.
Henceforth every vertex centers a conflict. Write
$s=s_4(G)$ and let $H=\{v:d_G(v)>4\}$. Then
\[
 |H|\le s,
 \qquad
 \sum_{v\in H}d_G(v)=s+4|H|\le5s.
\]

\paragraph*{Enlarging the exceptional set.}
In addition to the original exceptional components,
mark as exceptional every ordinary $K_3$ component
in $\widehat G$ whose corresponding edges have an
endpoint in $H$. This designation changes neither
$G$ nor $\widehat G$. Let $B'$ contain all host
endpoints of edges in the original or newly marked
exceptional components, and put $b'=|B'|$.

\begin{claim}\label{cl:exception-bound}
The enlarged set satisfies $b'\le b+16s$ and contains
every vertex of degree greater than four.
\end{claim}
\begin{claimproof}
If $a$ components are newly marked, then
$a\le\sum_{v\in H}d_G(v)\le5s$: each contains an
edge incident with $H$, and different conflict
components contain disjoint host-edge sets.
By locality, each such $K_3$ is represented by a
host triangle or a three-edge star. It has at most
four host endpoints, at least one in $H$.
The union of the endpoints of newly marked
components therefore has size at most $|H|+3a$,
giving
\begin{equation}\label{eq:majority-boundary}
 b'\le b+|H|+3a\le b+16s.
\end{equation}

Since no single-vertex cut is conflict-free, every
$v\in H$ centers a conflict. Its conflict component
is either originally exceptional or an ordinary
$K_3$ that is newly marked because its host edges
meet $H$. In either case, $v\in B'$, proving
$H\subseteq B'$.
\end{claimproof}

Classify the unmarked ordinary triangle components
of $\widehat G$ as in Theorem~\ref{thm:triangles}.
Let $T\subseteq V(G)$ consist of the vertices of
their rigid triangles, partitioned into rigid
classes as before. Let $S\subseteq V(G)$ consist
of the centers of their three-edge stars.
Every host endpoint of an unmarked $K_3$ has degree
at most four. Consequently $T\cap S=\emptyset$,
since a rigid triangle and a distinct centered
star would use at least five incident edges.
No vertex centers two unmarked stars, since these
would use six incident edges.

Every vertex outside $B'$ belongs to $T\cup S$.
Indeed, the conflict blocking its singleton cut
belongs to an unmarked $K_3$: either a rigid
triangle contains the vertex, or a three-edge
star is centered there. Orient the unmarked
star edges towards their centers.
Every vertex of $S$ has indegree three and
outdegree at most one, since its star supplies
three incoming edges and its host degree is
at most four.

Every exceptional conflict has all its host
endpoints in $B'$, so it can be checked once
$B'$ has been colored. Vertices in
$B'\setminus(T\cup S)$ may have unbounded degree;
their colors will be fixed before extending the
coloring to the remaining vertices.
Figure~\ref{fig:majority-sets} illustrates these sets.

\begin{figure}[tb]
\centering
\begin{tikzpicture}[x=1cm,y=1cm,font=\small,
  every node/.style={align=center,inner sep=2pt}]

\fill[blue!8] (0,0) rectangle (4.6,4.2);
\fill[teal!8] (8.2,0) rectangle (12.8,4.2);
\fill[orange,opacity=.10] (3.2,-.30) rectangle (9.6,4.50);

\draw[blue!60!black] (0,0) rectangle (4.6,4.2);
\draw[teal!60!black] (8.2,0) rectangle (12.8,4.2);
\draw[orange!70!black,line width=.65pt] (3.2,-.30) rectangle (9.6,4.50);

\fill[violet,opacity=.09] (6.4,1.55)
  ellipse [x radius=2.75,y radius=1.15];
\draw[violet!70!black,dashed] (6.4,1.55)
  ellipse [x radius=2.75,y radius=1.15];

\fill[red,opacity=.09] (6.4,3.05)
  ellipse [x radius=1.05,y radius=.70];
\draw[red!60!black] (6.4,3.05)
  ellipse [x radius=1.05,y radius=.70];

\node at (2.3,4.48) {$T$};
\node at (6.4,4.78) {$B'$};
\node at (10.5,4.48) {$S$};
\node at (6.4,1.40) {$B$\\original exceptional\\endpoints};
\node at (6.4,3.24) {$H$\\degree $>4$};

\node at (1.6,2.45)
  {$T\setminus B'$\\rigid vertices};
\node at (3.9,3.35)
  {$T\cap B'$\\rigid\\vertices};
\node at (8.9,3.35)
  {$S\cap B'$\\pinned\\centers};
\node at (11.2,2.45)
  {$S\setminus B'$\\free majority\\centers};

\node[anchor=north west] at (0,-.10) {$V(G)$};
\end{tikzpicture}
\caption{The host-vertex sets after singleton
preprocessing and marking.
The sets $B$ and $B'$ contain the host endpoints
of the original and enlarged exceptional
components, respectively. The sets $T$ and $S$
contain the rigid-triangle vertices and star
centers of the unmarked components.
We have $B\subseteq B'$, $T\cap S=\emptyset$,
$V(G)=B'\cup T\cup S$, and
$H\subseteq B'\setminus(T\cup S)$.
Centers in $S\cap B'$ have fixed colors during
each extension attempt.
The overlaps are schematic and need not be nonempty.}
\label{fig:majority-sets}
\end{figure}
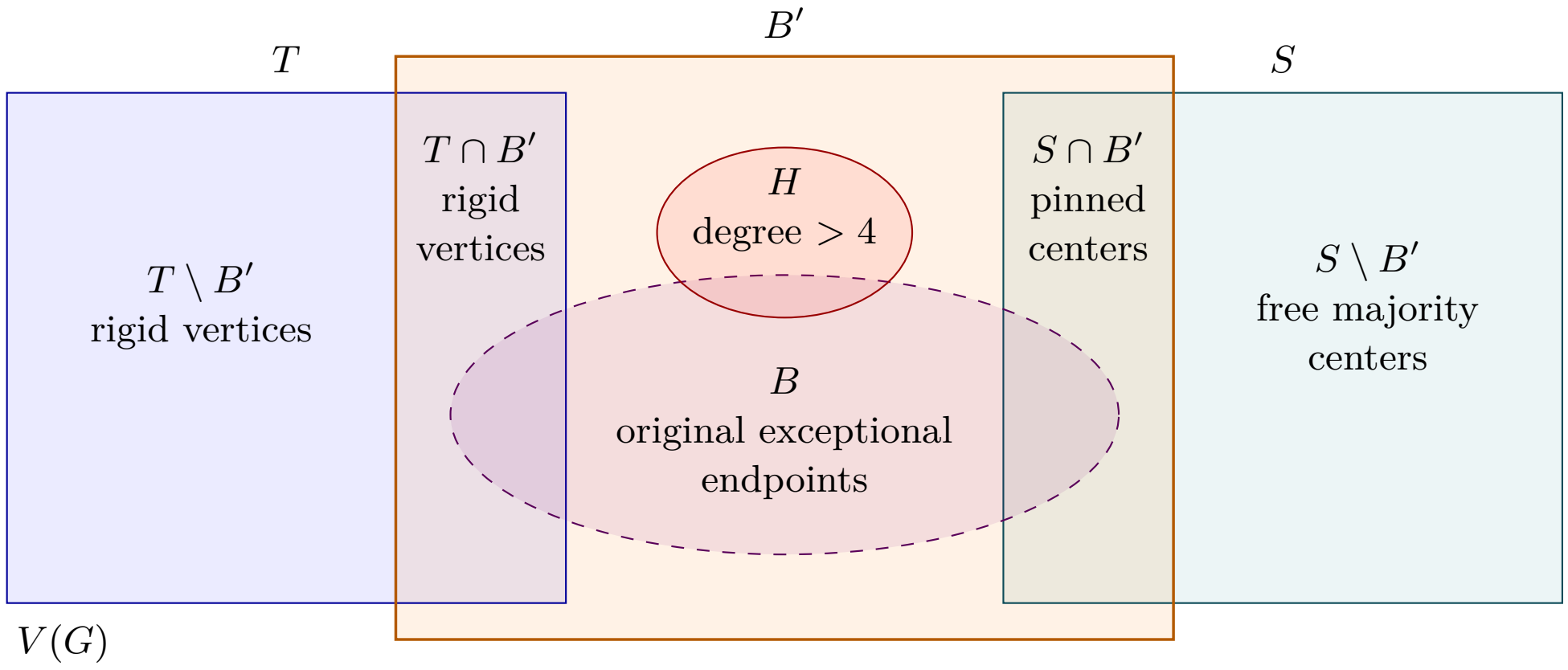

If $b'=0$, then $H=\emptyset$ and all components
are ordinary, so apply Theorem~\ref{thm:triangles}.
Henceforth assume $b'>0$.
In each extension attempt, the vertices of $B'$
receive prescribed colors.
We call the centers in $S\cap B'$ \emph{pinned}:
their colors remain fixed, but their majority
equations must still be satisfied.
The centers in $S\setminus B'$ are \emph{free}.
A rigid class is \emph{untouched} if it avoids $B'$.

\paragraph*{Propagating rigid-class labels.}
\begin{claim}\label{cl:many-classes}
If there are more than $|S\cap B'|$ untouched
rigid classes, then a CF-cut can be constructed
in polynomial time.
\end{claim}
\begin{claimproof}
Let $A_1,\ldots,A_r$ be the untouched rigid classes, where $r > |S\cap B'|$.
Initially, assign label $i$ to every vertex of $A_i$,
for each $i\in\{1,\ldots,r\}$, and leave all other
vertices unlabelled.
Vertices of untouched rigid classes retain their
initial labels, and all other vertices of
$T\cup B'$ remain unlabelled throughout the process.
In particular, every vertex of $B'$ remains unlabelled,
including the pinned centers in $S\cap B'$.

Propagate all labels in a single process.
Repeatedly choose an unlabelled free center
$v\in S\setminus B'$ with at least two predecessors
carrying the same label $i$, and assign label $i$
to $v$. Labels are never changed.
The two predecessors that caused this assignment
retain label $i$. Since $v$ has only three
predecessors, it can never later have two
predecessors carrying a different label $j\ne i$.
Continue until no further label can be assigned.

When propagation terminates, each label
$i\in\{1,\ldots,r\}$ defines a coloring
\[
 f_i(v)=
 \begin{cases}
 1,&\text{if $v$ carries label $i$},\\
 0,&\text{otherwise}.
 \end{cases}
\]
Every free majority equation holds under $f_i$.
A center labelled $i$ has two predecessors with
that label, which remain labelled $i$.
A center with a different label has two predecessors
carrying that different label, so at most one
predecessor carries $i$.
An unlabelled free center also has at most one
predecessor labelled $i$, since otherwise
propagation would continue.

Every unmarked rigid triangle is monochromatic
under $f_i$. Every exceptional conflict is
satisfied because all of $B'$ receives zero.
Thus only majority equations at pinned centers
might fail. Now fix a pinned center $v$ and consider all
candidate colorings $f_i$, one for each untouched
rigid class $A_i$.
The vertex $v$ receives zero in every coloring $f_i$,
and its majority equation fails under $f_i$
precisely when at least two of its predecessors
carry label $i$.
Since each predecessor carries at most one label,
at most one label can occur on two or more of
the three predecessors.
Therefore $v$ can invalidate at most one
candidate coloring.

Since $r>|S\cap B'|$, the pinned centers cannot
invalidate all $r$ colorings. Hence some coloring
$f_i$ satisfies every constraint.
It is nonconstant: the class $A_i$ receives one
and the nonempty set $B'$ receives zero.
Propagation makes at most $n$ label assignments.
The untouched rigid classes are nonempty and
pairwise disjoint, so $r\le n$.
There are therefore at most $n$ colorings
$f_1,\ldots,f_r$ to check.
Each propagation step can be implemented by scanning
the free centers, and each coloring can be checked
in polynomial time.
Hence a suitable coloring, and therefore
a CF-cut, can be found in polynomial time.
\end{claimproof}

\paragraph*{Guessing colors when few classes remain.}
Suppose that Claim~\ref{cl:many-classes} does not
apply, and let $t$ be the total number of rigid
classes. Classes meeting $B'$ number at most
$|T\cap B'|$, and untouched classes number at
most $|S\cap B'|$. Therefore
\begin{equation}\label{eq:root-bound}
 t\le |T\cap B'|+|S\cap B'|\le b'.
\end{equation}
Enumerate all assignments of colors in $\bits$
to the rigid classes and the vertices of
$B'\setminus T$, assigning a common color to
all vertices of each rigid class.
This uses at most
$t+|B'\setminus T|\le2b'$ bits.
All exceptional conflicts are now fully determined;
reject any assignment violating one.
Only the free star centers in $S \setminus B'$ remain unassigned.

\begin{claim}\label{cl:majority-extension}
For a fixed guess of these colors, feasibility
can be decided in polynomial time, and a feasible
extension can be recovered if one exists.
\end{claim}
\begin{claimproof}
Consider the digraph induced by all centers $S$,
including pinned centers. It has outdegree at
most one. Each weak component is an in-tree
directed towards a sink, or a directed cycle
with in-trees directed towards its vertices.
All inputs from outside $S$ have fixed colors.

First process all vertices outside directed cycles,
evaluating each in-tree from its leaves towards
its sink or its attachment to a cycle.
Cycle vertices are processed in the next step. 
When a vertex is processed,
all three inputs to its majority equation are
known, so its color is uniquely determined.
If the vertex is pinned and the determined color
differs from its prescribed color, reject the
current guess. Otherwise, assign the determined
color to the vertex and continue.

For each directed cycle, all off-cycle inputs
have now been evaluated.
Try each of the two colors for one cycle vertex,
respecting any prescribed color there.
Follow the cycle in its directed order.
At each subsequent vertex, the color of its
cycle predecessor and its two off-cycle inputs
determine its color uniquely.
Check every prescribed color and, finally,
the majority equation at the first vertex.
If a check fails, discard this initial choice
and try the other one.
If neither choice succeeds, reject the current
guess; otherwise, retain one successful coloring
of the cycle.

The components can be solved independently,
since every input from outside $S$ is fixed.
The tree colors are forced, and the two initial
choices on each cycle exhaust its possible
extensions. Thus the procedure decides feasibility
and constructs an extension whenever one exists.
It takes linear time in the dependency digraph,
after the exceptional conflicts have been checked.
\end{claimproof}

\paragraph*{Nonconstancy and running time.}
It suffices to test guesses using both colors
among the prescribed vertices.
Indeed, if every prescribed color equals $c$,
evaluation forces every off-cycle center to
have color $c$. Each cycle vertex then has two
off-cycle inputs of color $c$, forcing its color
to be $c$ as well. Thus such a guess has only
the constant extension.
A guess using both colors remains nonconstant
under every extension.

If there are more than $|S\cap B'|$ untouched
rigid classes, Claim~\ref{cl:many-classes}
constructs a CF-cut in polynomial time.
Otherwise, every nonconstant feasible coloring
determines a guess using both colors, and our
enumeration includes all such guesses.
Claim~\ref{cl:majority-extension} tests each guess
and constructs a feasible extension whenever one
exists. Since both colors are already prescribed,
any such extension gives a CF-cut.

There are at most
\[
 2^{t+|B'\setminus T|}
 \le 2^{2b'}
 \le 2^{2b+32s}
\]
guesses, each processed in polynomial time.
This proves the claimed constructive running time.
\end{proof}

\begin{table}[tb]
\caption{Parameterized extensions for local conflicts on arbitrary hosts. Isolated conflict vertices are permitted. $t_2,t_3,t_4$ count conflict components $K_2,K_3,C_4$, respectively; $C_4$ denotes a cycle, not a clique. Setting $s_4=0$ gives the maximum-degree-four specializations.}
\label{tab:component-parameters}
\centering
\begin{tabular}{>{\raggedright\arraybackslash}p{.25\linewidth}>{\raggedright\arraybackslash}p{.36\linewidth}>{\raggedright\arraybackslash}p{.29\linewidth}}
\toprule
Conflict components & Running time / parameter & Reference\\
\midrule
$K_1,K_3$ & $2^{O(s_4)}n^{O(1)}$ & Corollary~\ref{cor:component-fpt} (i)\\
$K_1,K_2,K_3$ & $2^{O(s_4+t_2)}n^{O(1)}$ & Corollary~\ref{cor:component-fpt} (i)\\
$K_1,K_3,C_4$ & $2^{O(s_4+t_4)}n^{O(1)}$ & Corollary~\ref{cor:component-fpt} (i) \\
$K_1,K_2,K_3$ & FPT in $s_4+\min\{t_2,t_3\}$ & Corollary~\ref{cor:component-fpt} (ii)\\
\bottomrule
\end{tabular}
\end{table}

\section{Proofs for locality and host structure}
\label{app:locality}

We first prove the scope-transversal algorithm,
then give the constructions for the component-order
hardness results and the two-vertex and singleton
deletion-set cases.
We then prove the clique hardness result,
describe the exactification construction,
and establish hardness with a supplied
universal vertex.
We conclude by summarizing the supplementary
boundaries and their consequences for
structural parameters.

\subsection{Proof of Theorem~\ref{thm:scope}}
\label{app:scope}

\begin{proof}[Proof of Theorem~\ref{thm:scope}]
Let $X$ be a supplied scope transversal.
For each of its $2^{|X|}$ colorings, we test
whether it extends to a nonconstant feasible
coloring of $G$.

Every conflict has a vertex in $X$, so its
conditioned relation involves at most two unknown
colors. Express this relation as a 2-CNF formula
by including, for each forbidden assignment,
the clause excluding that assignment.
One unknown variable gives unit clauses;
a violated constraint with no unknown variables
rejects the current guess.
Together, these formulas give a 2-SAT instance
of polynomial size, solvable in linear time
in its size~\cite{APT1979}.

To ensure nonconstancy, if the coloring of $X$
already uses both colors, any satisfying extension
suffices. If $X\ne\emptyset$ is monochromatic,
try each vertex outside $X$ fixed to the opposite
color. If $X=\emptyset$, fix an arbitrary vertex
to zero and try each other vertex fixed to one.
Complementation justifies fixing the first vertex
to zero. These tests cover every possible
nonconstant solution and add only a polynomial
factor.

To find a scope transversal, choose an uncovered
scope and branch on including one of its three
vertices. With budget $k$, this gives at most
$3^k$ leaves. Increase the budget until a
transversal is found, then apply the coloring
algorithm. The search and coloring costs add,
giving $O^*(3^{\tau_{\rm sc}})$ time.

By Observation~\ref{obs:coc-scope}, a supplied
$2$-COC can be used directly.
To find one, choose three vertices inducing
a connected subgraph in a component of order
at least three, and branch on deleting one of
them. Every $2$-COC must delete at least one
of these vertices. Increasing the budget until
a solution is found gives a minimum $2$-COC
in $O^*(3^{\coc_2(G)})$ time.
Applying the coloring algorithm completes
the proof.
\end{proof}

\subsection{Proof of Theorem~\ref{thm:coc-hardness}(i)}
\label{app:coc3}

Write $\fvs(G)$ for the minimum size of a vertex
set whose deletion leaves a forest, and
$\tw(G),\pw(G)$ for treewidth and pathwidth.
Treedepth $\td(G)$ is the minimum height of a
rooted forest whose ancestor closure contains $G$,
counting height in vertices.
Vertex integrity is
$\vi(G)=\min_X(|X|+\max_K|V(K)|)$,
where $K$ ranges over the components of $G-X$
and the maximum is zero if none remain.

The construction for part~(i) also satisfies
\[
  \coc_3(G)\le6,\quad \fvs(G)\le6,\quad
  \tw(G),\pw(G)\le7,\quad
  \td(G)\le8,\quad \vi(G)\le9.
 \]

\begin{proof}[Proof of Theorem~\ref{thm:coc-hardness}(i)]
We reduce from $\operatorname{SAT}_c(R)$, which is NP-complete by Theorem~\ref{thm:SATR}. We may ensure that there is at least one constraint by appending $R(0,0,0)$. Unused variables can be discarded in this source problem.

\paragraph*{The two constant graphs.}
Create two disjoint 5-cliques $P=\{p_0,\ldots,p_4\}$ and $Q=\{q_0,\ldots,q_4\}$. For the edges of $P$, put the five conflict pairs
\begin{equation}\label{eq:anchor}
 \{p_ip_{i-1},p_ip_{i-2}\}\qquad(i=0,\ldots,4),
\end{equation}
where subscripts are modulo five. Use the analogous pairs in $Q$.

\begin{claim}\label{cl:anchor-rigid}
These pairs partition the edges of each clique. Every feasible coloring is constant on each clique.
\end{claim}
\begin{claimproof}
Each of the ten edges occurs once. Orient the two edges of the pair at $p_i$ into $p_i$. Feasibility requires every vertex to have a same-colored in-neighbor. In a nonconstant coloring of five vertices, a smaller color class has one or two vertices. A singleton has no in-neighbor of its color. For two vertices, their one connecting edge is oriented in just one direction, so both cannot have an in-neighbor within that class. This proves rigidity, and the argument for $Q$ is identical.
\end{claimproof}

\paragraph*{Occurrences and clause paths.}
For each constraint $C=R(a,b,c)$, create three distinct
occurrence vertices $a_C,b_C,c_C$, one for each position
in the constraint. These vertices are distinct even
when two positions contain the same variable or constant. Add the path $a_Cb_Cc_C$ and make its two edges a conflict pair. Every vertex $x$ outside $P\cup Q$, including each auxiliary introduced below, also receives the edges $xp_0,xq_0$, which form a conflict pair.

If the two cliques have the same color, these latter pairs force all outside vertices to have that color. Hence every nonconstant solution has opposite colors on $P$ and $Q$. By complementing colors if necessary, we interpret their colors as zero and one, respectively. Since $p_0$ and $q_0$ have opposite colors, the conflict
$\{xp_0,xq_0\}$ is satisfied for either color of $x$.

\paragraph*{Equalizing variable occurrences.}
For each variable $x$, 
list the vertices representing
its $t$ occurrences 
as $x_1,\ldots,x_t$ in a cyclic order, with $x_{t+1}=x_1$. 
These are the occurrence vertices already introduced
in the clause paths. For example, if the $i$th
occurrence of $x$ occupies the first position in
$C=R(a,b,c)$, then $x_i=a_C$.
Add $t$ fresh auxiliary vertices $z_1,\ldots,z_t$,
where $z_i$ is used to synchronize $x_i$ and $x_{i+1}$.
For each $i$, add the edges $z_ip_0,z_iq_0$ as a
conflict pair, together with the following four
conflict pairs:
\begin{equation}\label{eq:sync}
\begin{split}
 \{p_1x_i,p_1z_i\},\quad \{q_1x_i,q_1z_i\},\\
 \{p_2z_i,p_2x_{i+1}\},\quad \{q_2z_i,q_2x_{i+1}\}.
\end{split}
\end{equation}
The above notation also works when $t=1$, where there is only one occurrence vertex $x_1$ and one auxiliary vertex $z_1$.
A constant-zero occurrence $x$ instead receives the pair $\{xp_1,xp_2\}$. A constant-one occurrence receives $\{xq_1,xq_2\}$. Constants are not included in variable synchronization.

\begin{claim}\label{cl:anchor-sync}
When $P$ has color zero and $Q$ has color one, the pairs in Equation~\eqref{eq:sync} force all occurrences of each variable to agree. The constant-occurrence pairs force the indicated constant values.
\end{claim}
\begin{claimproof}
Let $f$ be a feasible coloring with color zero on $P$
and color one on $Q$.
The first row of Equation~\eqref{eq:sync} forbids
$f(x_i)=f(z_i)=1$ and $f(x_i)=f(z_i)=0$.
Thus $f(x_i)\ne f(z_i)$.
The second row similarly forces
$f(z_i)\ne f(x_{i+1})$, giving
$f(x_i)=f(x_{i+1})$.
Each auxiliary vertex therefore has the opposite
color from the occurrence vertices of its variable.
For a constant occurrence, both neighbors in its
pair have the indicated constant color.
If the occurrence vertex had the other color,
both edges would cross.
\end{claimproof}

\paragraph*{Verifying the restrictions.}
The internal clique edges and clause-path edges are
distinct. Every other edge joins an occurrence or
auxiliary vertex to one of the two cliques.
Each such vertex has an edge to each of $p_0,q_0$,
and these two edges form a conflict pair.
Each variable-occurrence vertex and each
auxiliary vertex also has exactly one edge to each of
$p_1,p_2,q_1,q_2$, paired as specified in
Equation~\eqref{eq:sync}.
A constant-occurrence vertex instead has two additional
edges: to $p_1,p_2$ for constant zero, or to $q_1,q_2$
for constant one. These form its constant-forcing pair.
The distinct clique vertices ensure that no edge is
reused, even when $t=1$.
Thus every edge belongs to exactly one conflict pair,
and every conflict is local.
Fresh occurrence and auxiliary vertices ensure simplicity. Since every outside vertex is adjacent to both $p_0$ and $q_0$, and there is at least one constraint, the graph is connected.

Delete the set
\begin{equation}\label{eq:sixset}
 X=\{p_0,p_1,p_2,q_0,q_1,q_2\}.
\end{equation}
The components that remain are the clause paths,
the isolated auxiliary vertices $z_i$, and the
edges $p_3p_4,q_3q_4$. This proves the $3$-COC and feedback-vertex bounds. Adding $X$ to every bag of a width-one path decomposition of $G-X$ gives pathwidth at most seven, and therefore treewidth at most seven. Placing the six vertices of $X$ in a chain above depth-two rooted forests for $G-X$ gives treedepth at most eight. The set $X$ also gives vertex integrity at most $6+3=9$.

\paragraph*{Correctness of the reduction.}
Given a satisfying source assignment of the $\operatorname{SAT}_c(R)$ instance, construct a
coloring $f$ of $G$ as follows. Color every vertex
of $P$ zero and every vertex of $Q$ one.
Give each occurrence vertex the value of its source
variable or constant, and give each auxiliary vertex
$z_i$ the color $1-f(x_i)$.
All pairs hold by Claim~\ref{cl:anchor-sync} and the source constraints. The two cliques guarantee nonconstancy.

Conversely, let $f$ be a nonconstant feasible coloring
of the constructed instance.
Claim~\ref{cl:anchor-rigid} makes the cliques
monochromatic, and the pairs to $p_0,q_0$ force
them to have opposite colors.
By complementing all colors if necessary, assume
that every vertex of $P$ has color zero and every
vertex of $Q$ has color one.
Claim~\ref{cl:anchor-sync} recovers a value for
each variable and the correct values for constants. For each source constraint $C=R(a,b,c)$, the
clause-path conflict ensures
$R(f(a_C),f(b_C),f(c_C))$.
The recovered source assignment therefore satisfies
every constraint.

For $c$ constraints and $r$ variable occurrences, excluding constants, the graph has $10+3c+r$ vertices and $20+14c+8r$ edges. The reduction is polynomial, and the problem belongs to NP.
\end{proof}

\subsection{Proof of Theorem~\ref{thm:coc-hardness}(ii)}
\label{app:nonlocal2}

\begin{proof}[Proof of Theorem~\ref{thm:coc-hardness}(ii)]
Use the construction in Appendix~\ref{app:coc3}, changing only the clause occurrences. For a constraint $C=R(a,b,c)$, create distinct vertices $a_C,b_C^L,b_C^R,c_C$ and add the two edges $a_Cb_C^L,b_C^Rc_C$. Make these edges a conflict pair. The two middle occurrences represent the same source symbol $b$. If it is a variable, include both in its synchronization cycle; if it is a constant, give both the appropriate constant-occurrence pairs. Let $f$ be a nonconstant feasible coloring of the
constructed graph. The synchronization or
constant-occurrence pairs ensure that
$f(b_C^L)=f(b_C^R)$.
The new clause conflict requires
\[
 \bigl(f(a_C)=f(b_C^L)\bigr)
 \lor
 \bigl(f(b_C^R)=f(c_C)\bigr),
\]
which, since the two middle colors agree, is
equivalent to
\[
 R\bigl(f(a_C),f(b_C^L),f(c_C)\bigr).
\]
Under the corresponding source assignment, this
enforces the constraint $R(a,b,c)$.

Each occurrence vertex has its own edges to $p_0,q_0$,
which form a conflict pair.
A variable-occurrence vertex also receives the
conflict pairs involving auxiliary vertices, whereas
a constant-occurrence vertex receives the pair
forcing its constant value.
As in the preceding construction, every edge belongs
to exactly one conflict pair, and the graph remains
simple and connected. The same extension and restriction arguments as in Theorem~\ref{thm:coc-hardness}(i) prove correctness. Deleting $X$ now leaves only isolated vertices and edges. Only the new clause pairs are nonlocal. This proves the theorem.
\end{proof}

\subsection{Deletion sets of size two and one}
\label{app:twohubs}

We first reduce the local $3$-COC hardness bound
to two by allowing edges to participate in several
conflicts. We then show that a supplied singleton
$a$-COC permits polynomial time for every fixed $a$,
even without locality.

\begin{theorem}\label{thm:twohubs}
With arbitrary local conflicts, \CF{} is NP-complete with a supplied $3$-COC of size two. The constructed hosts are interval graphs, have a universal vertex, and satisfy
\[
 \fvs(G)\le2,\quad \pw(G),\tw(G)\le3,\quad \td(G)\le4,\quad \vi(G)\le5.
\]
\end{theorem}
\begin{proof}
We reduce from $\operatorname{SAT}_c(R)$, whose hardness
follows from Theorem~\ref{thm:SATR}.
Create adjacent vertices $p,q$.
For each source constraint $C=R(a,b,c)$, create three
distinct occurrence vertices $a_C,b_C,c_C$.
Join every vertex $u\notin\{p,q\}$, including each
auxiliary introduced below, to both $p$ and $q$,
and make $\{up,uq\}$ a conflict pair.
In any feasible coloring $f$, if $f(p)=f(q)$,
these pairs force $f$ to be constant.
Thus, in a nonconstant feasible coloring, we may
assume $f(p)=0$ and $f(q)=1$ by complementing colors
if necessary.

For an occurrence vertex $u$ representing constant
zero, add the conflict $\{pq,pu\}$, which forces
$f(u)=0$. For constant one, use $\{pq,qu\}$,
which forces $f(u)=1$.
For distinct vertices $u,v\notin\{p,q\}$, the conflicts
\[
 \{pu,pv\},\qquad \{qu,qv\}
\]
together force $f(u)\ne f(v)$.

To force consecutive occurrences $u,v$ of the
same variable to agree, introduce a fresh
auxiliary vertex $w$ and add the conflicts
$\{pu,pw\}$, $\{qu,qw\}$, $\{pw,pv\}$,
and $\{qw,qv\}$.
They enforce
$f(u)\ne f(w)$ and $f(w)\ne f(v)$, hence $f(u)=f(v)$.
Apply this to every consecutive pair in an ordering
of each variable's occurrence vertices.

Finally, for each constraint $C=R(a,b,c)$, add the
path $a_Cb_Cc_C$ and make its two edges a conflict
pair. This enforces $R(f(a_C),f(b_C),f(c_C))$.

All conflicts are local, although an edge may
participate in several conflicts.
Deleting $p,q$ leaves the clause paths as separate
$P_3$ components and every auxiliary vertex isolated.

Given a satisfying source assignment, set
$f(p)=0$ and $f(q)=1$, and give each occurrence
vertex its source value.
For each auxiliary $w$ introduced between
occurrences $u,v$ of the same variable, set
$f(w)=1-f(u)$.
The resulting coloring is feasible and nonconstant.
Conversely, any nonconstant feasible coloring,
after complementation if necessary, has
$f(p)=0$ and $f(q)=1$.
The added conflicts force constant occurrences
to have their prescribed values and occurrences
of the same variable to have equal colors.
Each clause-path conflict enforces its source
constraint. Thus these colors give a satisfying
source assignment.
The construction is polynomial, and a nonconstant
feasible coloring can be checked in polynomial time,
proving NP-completeness.

Both $p$ and $q$ are universal vertices.
The graph $G-\{p,q\}$ is a forest with pathwidth
at most one and treedepth at most two.
Adding $p,q$ to every bag gives pathwidth at most
three, and hence treewidth at most three.
Placing $p,q$ in a chain above a rooted forest
representing $G-\{p,q\}$ gives treedepth at most four.
The same deletion set gives feedback-vertex number
at most two and vertex integrity at most $2+3=5$.
For an interval representation, represent each
component of $G-\{p,q\}$ by intervals contained
in a separate segment of the real line.
Represent $p$ and $q$ by two intervals spanning
all these segments.
The resulting intersection graph is exactly $G$.
\end{proof}

\begin{proposition}[One-vertex deletion set]\label{prop:onevertex}
For any fixed $a$, \CF{} with arbitrary, possibly nonlocal conflicts can be solved in $O(n2^an^{O(1)})$ time when supplied with a singleton $a$-COC $\{p\}$.
\end{proposition}
\begin{proof}
Let $(G,\widehat G)$ be the input instance, and let
$\{p\}$ be the supplied $a$-COC.
Thus every component of $G-p$ has at most $a$ vertices.
By complement symmetry, we may restrict attention
to colorings with $f(p)=0$. If a nonconstant feasible coloring exists, one component $K$ of $G-p$ contains a vertex of color one. Keep its colors and reset every other outside component to zero. There are no edges between distinct components, so the new cut is a subset of the old cut and remains conflict-free. Enumerate the choice of $K$ and all nonzero colorings of its at most $a$ vertices, setting everything else to zero and checking all conflicts.
\end{proof}

\subsection{Hardness on cliques}
\label{app:cliques}

Adding missing host edges without assigning them
conflicts preserves every feasible vertex coloring.
This transfers the local degree-five hardness
to cliques.

\begin{proposition}\label{prop:clique}
\CF{} is NP-complete on cliques with local matching conflicts, allowing unpaired edges. Consequently the same holds on split graphs and interval graphs under this conflict regime.
\end{proposition}
\begin{proof}
Take an instance from Theorem~\ref{thm:degree-five}(i) and add every missing edge of $G$, giving it no conflicts. 
This leaves all conflict constraints and the
nonconstancy condition unchanged. 
The resulting clique has exactly the same feasible colorings. All old conflicts stay local and form a matching; the new edges are unpaired. Cliques are both split graphs and interval graphs.
\end{proof}

\subsection{Making every edge participate in a conflict}
\label{app:exactify}

The following construction removes unpaired edges
while preserving locality and feasible colorings,
but need not preserve the host graph class.

\begin{proposition}\label{prop:exactify}
Any local matching-conflict instance can be transformed in polynomial time into an equivalent local exactly $1$-regular instance.
\end{proposition}
\begin{proof}

Let $U$ be the rigid octahedron with the conflict
pairs specified in Equation~\eqref{eq:rigidU}.
For each unpaired edge $xy$, attach a fresh copy
of $U$ by identifying its vertex $0$ with $x$.
Retain its internal pairs, add its missing edge $x3$, and pair $xy$ with $x3$. Rigidity forces the new block to the color of $x$, so $x3$ is always monochromatic. Thus the new conflict imposes no
restriction on colorings of the original instance. All new vertices agree with an old vertex, preserving nonconstancy. Every old unpaired edge and every new edge now has exactly one local mate. The construction does not preserve a clique, degree bound, or small COC in general.
\end{proof}

\subsection{Hardness with a universal vertex}
\label{app:domination}

We add a universal vertex to a hard local
degree-five instance and use rigid octahedra
to give every added edge exactly one
conflicting partner.

\begin{theorem}\label{thm:domination}
\CF{} is NP-complete with local exactly $1$-regular conflicts and a supplied universal vertex. All other vertices can have degree at most six.
\end{theorem}
\begin{proof}
Take a hard connected instance $H$ from Theorem~\ref{thm:degree-five}(i) and distinct vertices $a,b$ of $H$. Add a vertex $u$ adjacent to every old vertex and pair $ua$ with $ub$. For each other old vertex $v \in V(H) \setminus \{a, b\} $, attach a fresh rigid $U_v$ with vertex zero identified with $u$, add its missing edge $u3_v$, and pair $uv$ with $u3_v$. Keep every internal pair of every block. All new blocks must agree with $u$, so their added edges $u3_v$ never cross. The vertex $u$ is universal, including to vertex $3_v$ of each block by the added edge. Every edge has exactly one local mate.

A nonconstant solution on $H$ extends by coloring $u$ like $a$ and every block like $u$. The pair $\{ua,ub\}$ holds because $ua$ does not cross. Conversely, if every old vertex has the same color, that special pair forces $u$ to have it too, and rigidity makes the entire target constant. Thus a nonconstant target solution restricts to a nonconstant solution on $H$. Old vertices gain one edge, so have degree at most six. New block vertices have degree four, except $3_v$, whose degree is five. Only $u$ can have unbounded degree. Polynomial size and NP membership complete the proof.
\end{proof}

\subsection{Further structural boundaries}
\label{app:further-boundaries}

Table~\ref{tab:further-boundaries} summarizes
the supplementary results established in the
preceding subsections.
Together with Theorem~\ref{thm:coc-hardness},
these results give hardness for several
structural parameters.

\begin{table}[tb]
\caption{Further boundaries for \CF{}.
``Matching'' permits unpaired edges.
The deletion sets specified below are supplied
with the input.}
\label{tab:further-boundaries}
\centering
\begin{tabular}{
@{}
>{\raggedright\arraybackslash}p{.42\linewidth}
>{\raggedright\arraybackslash}p{.25\linewidth}
>{\raggedright\arraybackslash}p{.25\linewidth}
@{}}
\toprule
Restriction on $G$
& Restriction on conflicts
& Complexity and proof\\
\midrule

Clique
& Local, matching
& NP-complete
(Appendix~\ref{app:cliques})\\
\addlinespace

A supplied universal vertex; every other
vertex has degree at most six
& Local, exactly $1$-regular
& NP-complete
(Appendix~\ref{app:domination})\\
\addlinespace

Interval graph with a universal vertex
and a supplied $3$-COC of size two
& Local; an edge may participate in
several conflicts
& NP-complete
(Appendix~\ref{app:twohubs})\\
\addlinespace

A supplied singleton $a$-COC
& Arbitrary, possibly nonlocal
& $2^a n^{O(1)}$;
polynomial for fixed $a$
(Appendix~\ref{app:twohubs})\\

\bottomrule
\end{tabular}
\end{table}

The interval graphs in the third row also satisfy
\[
 \fvs(G)\le2,\qquad
 \tw(G),\pw(G)\le3,\qquad
 \td(G)\le4,\qquad
 \vi(G)\le5.
\]
Thus local $3$-COC hardness occurs already at
deletion-set size two when edges may participate
in several conflicts, whereas size one is
polynomial even without locality.

The clique result allows unpaired edges.
Proposition~\ref{prop:exactify} transforms local
matching conflicts into local exactly
$1$-regular conflicts, but does not preserve
the clique condition. Hardness on cliques
with exactly $1$-regular conflicts therefore
remains open.

\begin{corollary}[Hardness for structural parameters]
\label{cor:parameter-hardness}
Unless $\mathrm P=\mathrm{NP}$, \CF{} admits
no XP algorithm, and hence no FPT algorithm,
parameterized by any one of the following:
\begin{enumerate}[(i)]
\item with local exactly $1$-regular conflicts:
$\coc_3(G)$, feedback-vertex number, treewidth,
pathwidth, treedepth, vertex integrity,
or domination number;

\item with local matching conflicts, allowing
unpaired edges:
independence number, neighborhood diversity,
clique-width, rank-width, or distance to
cluster graphs;

\item with exactly $1$-regular conflicts,
without requiring locality:
$\coc_2(G)$.
\end{enumerate}
\end{corollary}

\begin{proof}
The construction in Appendix~\ref{app:coc3}
establishes NP-hardness with constant bounds
on all parameters in (i) except domination number.
Theorem~\ref{thm:domination} supplies that case
with domination number one.
For (ii), Proposition~\ref{prop:clique} gives
hardness on cliques, where every listed parameter
is bounded by a constant.
Part (iii) follows from
Theorem~\ref{thm:coc-hardness}(ii).
An XP algorithm for any of these parameters
would therefore solve the corresponding
NP-complete restriction in polynomial time.
\end{proof}

\label{page:appendixend}
\end{document}